\documentclass[conference]{IEEEtran}
\IEEEoverridecommandlockouts
\usepackage{cite}
\usepackage{amsmath,amssymb,amsfonts}
\usepackage{algorithmic}
\usepackage{graphicx}
\usepackage{textcomp}
\usepackage{xcolor}
\begin{document}

\title{Machine Learning for Encrypted Malicious Traffic Detection: Approaches, Datasets and Comparative Study\\}

\author{\IEEEauthorblockN{Zihao Wang, Kar-Wai Fok, Vrizlynn L. L. Thing}
\IEEEauthorblockA{\textit{Cybersecurity Strategic Technology Centre} \\
\textit{ST Engineering}\\
Singapore \\
\{zihao.wang, fok.karwai\}@stengg.com, vriz@ieee.org}
}
\maketitle

\begin{abstract}
As people’s demand for personal privacy and data security becomes a priority, encrypted traffic has become mainstream in the cyber world. However, traffic encryption is also shielding malicious and illegal traffic introduced by adversaries, from being detected. This is especially so in the post-COVID-19 environment where malicious traffic encryption is growing rapidly. Common security solutions that rely on plain payload content analysis such as deep packet inspection are rendered useless. Thus, machine learning based approaches have become an important direction for encrypted malicious traffic detection. In this paper, we formulate a universal framework of machine learning based encrypted malicious traffic detection techniques and provided a systematic review. Furthermore, current research adopts different datasets to train their models due to the lack of well-recognized datasets and feature sets. As a result, their model performance cannot be compared and analyzed reliably. Therefore, in this paper, we analyse, process and combine datasets from 5 different sources to generate a comprehensive and fair dataset to aid future research in this field. On this basis, we also implement and compare 10 encrypted malicious traffic detection algorithms. We then discuss challenges and propose future directions of research.

\end{abstract}


\section{\bf Introduction}

In recent years, with the increasing demand on privacy and data security, enterprises are choosing to use encryption mechanisms to protect their application traffic transmission. In view of this trend, encrypted traffic data volume has increased dramatically in the global communication network. Gartner has reported that more than 80\% of traffic in 2019 are encrypted [1]. The trend of increasing traffic encryption from 2014 to the beginning of 2021, from the Google Transparency Report, is shown in Fig. 1. The percentage of encrypted web traffic on the Internet has increased from around 50\% in 2014 to around 95\% after 2020. 97\% of the world’s top 100 sites are utilizing the secure hypertext transfer protocol (HTTPs) [2].

\begin{figure}
\centerline{\includegraphics[width=21pc]{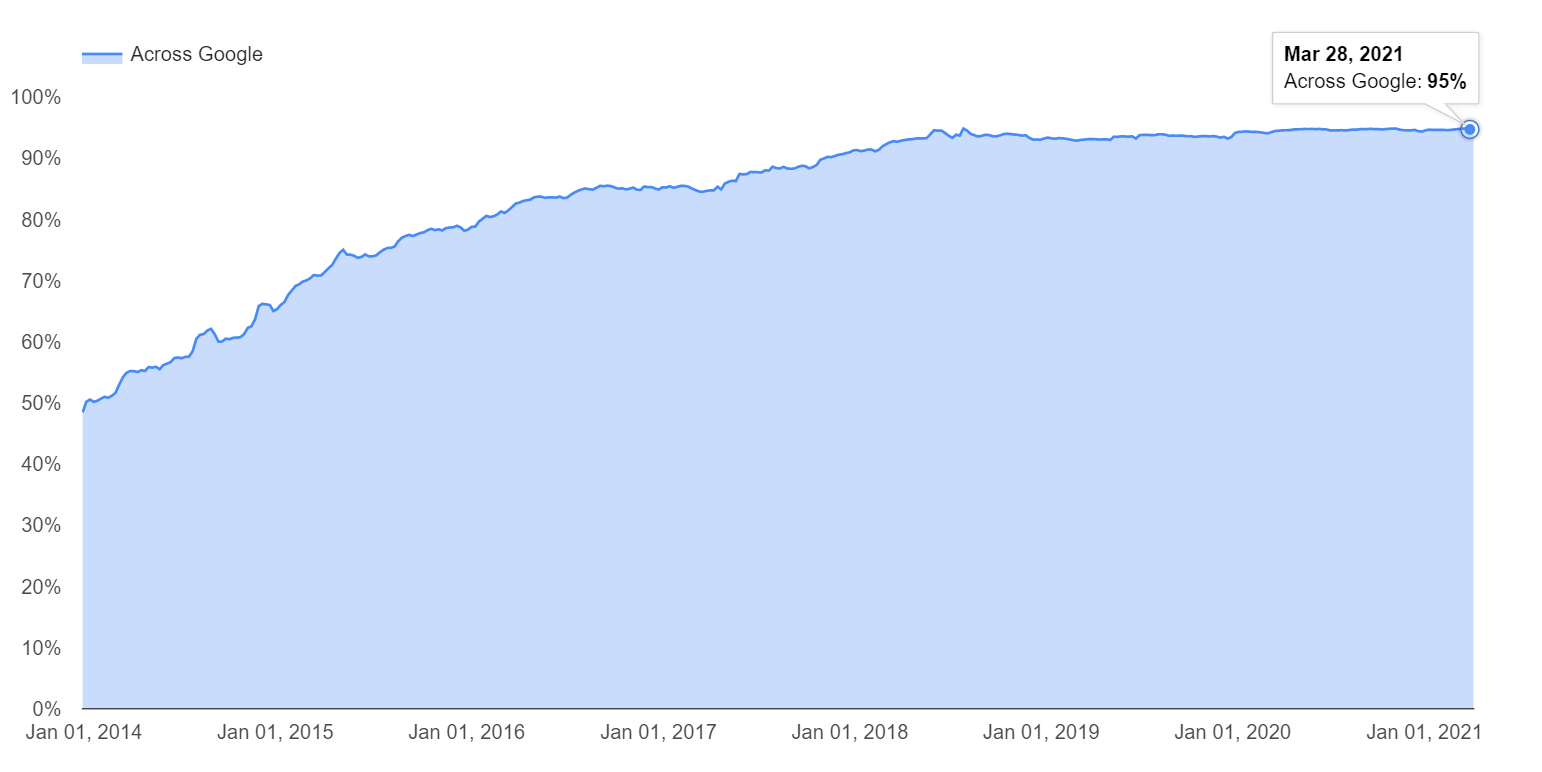}}
\caption{Trend of encrypted traffic across google [2]}
\end{figure}

While most people and enterprises enjoy the traffic privacy and data protection provided by encrypted traffic, adversaries are also leveraging encryption to evade detection of their malicious activities. With encrypted traffic, it poses a great challenge to antivirus software and firewalls which cannot decipher the traffic contents. In a 2018 cybersecurity report, Cisco noted that as of October 2017, about 280,000 of the 400,000 analyzed malware had communicated through encryption methods [3]. According to Zscaler’s 2020 Encrypted Attacks Report [4], there are more than 260\% increase in Secure Sockets Layer (SSL) based threats and more than 500\% increase in ransomware, where encrypted web traffic was utilized. Therefore, starting from the COVID-19 period, we are continuing to see accelerated growth in malicious traffic encryption. The report also pointed out that companies are at greater risk now because current cyber security systems cannot inspect 100\% of the network traffic. Therefore, it is timely that we investigate further on the detection of encrypted malicious traffic and the efficacy of existing research when dealing with this challenge.

The goal of this survey is to provide a comprehensive overview of machine learning based methods for encrypted malicious traffic detection. We also propose a framework to aid with the systematic discussion and analysis of machine learning based encrypted malicious traffic detection models. We also create a model training dataset that is composed of public traffic data from various sources, which is more comprehensive than the datasets used in existing research. This dataset is used to conduct fair comparative experiments on the different proposed feature sets and algorithms. Finally, the current challenges and future directions are discussed. Therefore, the contribution of this paper is:

\begin{itemize}
\item Review and discuss the strengths, limitations and compare the different techniques based on the machine learning framework we proposed. 
\item Conduct a relatively comprehensive collation and analysis of the traffic datasets currently available and analyse their characteristics, limitations and applicability. Such analysis enables the dataset to better serve this machine learning based approach.
\item Categorise the features of machine learning models, and discuss their strengths, limitations, applicable scenarios, and possible optimization directions.
\item Conduct comparative experiments based on the same training dataset, with different algorithms and feature sets to find a more reliable and consistent result. To the best of our knowledge, the dataset we built is the most comprehensive training dataset that is composed entirely from open public data sources.
\end{itemize}

This paper is organized into six sections. Section 2 presents the related surveys. In Section 3, we provide an introduction of the different detection techniques and describe a general framework of machine learning based encrypted malicious traffic detection. The strengths, limitations and comparison among the existing works based on our proposed framework are discussed in Section 4. In Section 5, we present the setup of our experiments and conduct the performance evaluations. We conclude the paper in Section 6, by discussing the remaining challenges and future directions.

\section{\bf RELATED SURVEYS}

In this section, we introduce related survey works. Velan et al. [6] provided us with detailed definitions and information of widely used traffic encryption protocols in their survey paper. It had also surveyed works on the classification of encrypted traffic before 2014. In contrast, we intend to conduct an in-depth study of existing research to analyse the various detection methods that have emerged in recent years. 

Conti et al. [56] conducted an in-depth survey for state of the art network traffic analysis generated by mobile devices. Three criteria are presented by them for a systematic classification of existing works, which are the objective of traffic analysis; the network traffic capture point and the targeted mobile platforms. Zhang et al. [61] bridged the gap between deep learning and mobile and wireless networks through a comprehensive review of the crossovers between these two areas. The authors reviewed almost 600 research papers and also discuss techniques and platforms which enhance the deployment of deep learning on mobile environments. Wang et al. [95] focused on the research of applying deep learning methods to encryption traffic classification of mobile service based on dataset selection, model input design and model architecture. The authors also proposed a general framework for deep learning based mobile encrypted traffic classification. Furthermore, they pinpointed some noteworthy problems and challenges of deep learning in encrypted traffic classification as well. Although [56][61][95] include works that are applicable to both encrypted traffic and unencrypted traffic, they do not pay much attention to malicious traffic detection and focus on mobile devices. In our review paper, we not only include works related to encrypted malicious traffic detection but also focus on platforms other than mobile devices.

Berman et al. [68] provided a comprehensive review for deep learning based cyber security applications including network traffic identification, network intrusion detection, malware classification and several others. Furthermore, the authors also provide a short introduction of each deep learning algorithm, such as restricted boltzmann machines, recurrent neural network (RNN) and generative adversarial network (GAN), and cover a wide range of attack types, such as malware, botnets and spam. Abbasi et al [59] presented a review of deep learning methods for network traffic monitoring and analysis(NTMA). Same as [68] The authors also provide us with detailed definitions and fundamental background of deep learning algorithms, such as Multi-layer perceptron (MLP), Convolutional neural networks (CNN), Long short term memory (LSTM), Auto-encoder (AE) and GAN models. Aceto et al. [98] presented an overview of the key subjects of traffic analysis where deep learning is expected to be attractive. The authors also provide a systematic taxonomy and categorize the existing deep learning based traffic classifications. Then, they proposed a general deep learning based framework for encrypted and mobile traffic classification with a rigorous definition of milestone(traffic object; type(s) of input data; tasks of classification and deep learning architecture). Zhai et al. [97] proposed a ‘six-step method’ framework for encrypted malicious traffic detection. The authors also reviewed the existing deep learning based encrypted malicious traffic detection approaches based on the proposed framework. 20 existing public datasets are also sorted and discussed based on their applicable scenarios with pros and cons. In contrast, our study provides a more comprehensive review of not only deep learning based approaches, but also traditional machine learning based approaches and Detailed feature analysis.

\section{\bf PRELIMINARIES}

Many mature technologies and methods for the detection and classification of unencrypted traffic, such as payload based deep packet inspection (DPI) methods and port-based identification methods, are already in existence. However, with traffic encryption mechanisms, these traditional methods are no longer applicable. Moreover, DPI methods had drawn concerns over user data privacy. On the other hand, port-based traffic identification methods assume that applications use well-known Transmission Control Protocol/User Datagram Protocol (TCP/UDP) port numbers assigned by Internet Assigned Numbers Authority (IANA). Therefore, once applications do not follow the standards of IANA, such as applying dynamic ports or encryption protocols like Peer to Peer (P2P) protocol, the port-based traffic identification methods can no longer identify the traffic and applications [16].

To perform traffic encryption, there is currently a variety of mechanisms, such as SSL, Transport Layer Security (TLS), Virtual Private Network (VPN), Secure Shell Protocol (SSH), and P2P. These encryption algorithms work differently; some at the transport layer while others at the application layer, making encrypted traffic classification a challenging task [5]. Even with the same encryption mechanism, the encrypted traffic presents varying data distribution characteristics due to the different distribution and utilization of the original traffic [15]. Thus, most of the research focuses on the binary classification of encrypted traffic, which is to identify malicious traffic among legitimate traffic.

Flow based machine learning methods have been observed to be the most common approach for encrypted traffic classification, such as [11-17][31-33][75-77]. However, it is worth noting that the collection of training datasets and the selection of features for encrypted traffic detection remains an area of vigorous research.

In recent years, researchers have proposed many detection methods based on machine learning, which can be generally classified into traditional machine learning and deep learning. Traditional Machine Learning can be further classified into two sub-groups: supervised learning and unsupervised learning. In the field of supervised learning, Shekhawat et al. [29] applied three supervised learning algorithms (Random Forest (RF), Support Vector Machine (SVM), and XGBoost) to the problem of distinguishing HTTPs malicious traffic and HTTPs legitimate traffic. Stergiopoulos et al. [14] conducted comparative experiments by using seven different supervised learning algorithms such as k-nearest neighbors (KNN), Classification And Regression Trees (CART), and Naïve Bayes to detect malicious traffic from a dataset with more than one encryption protocol. Ma et al. [32] proposed an enhanced KNN algorithm to train an encrypted traffic detection model, which enhances the KNN distance calculation. For unsupervised Learning, Chen et al. [13] proposed an improved density peaks clustering algorithm to enhance the accuracy and efficiency of encrypted malicious traffic detection. Celik et al. [17] compared the performance of K-means, one-class support vector machine (OCSVM), least squares anomaly detection (LSAD), and KNN algorithms by using tamper resistant features, such as Goodput and ratio between maximum packet over minimum packet. Zhang et al. [93] proposed a novel clustering algorithm for identifying encrypted traffic by applying harmonic mean to clustering distance metrics. As for the research on deep learning in encrypted traffic detection, most researchers focus on studying the performance of CNN, RNN, and AE. Bazuhair et al. [11] proposed a new encoding method where TLS/SSL connection features are converted into images. Such an image dataset is utilized to train the CNN model. Yao et al. [15] proposed an RNN with attention mechanism for encrypted traffic classification. Prasse et al. [25] applied a combination of neural domain name features and numeric flow features as the feature set for RNN, and showed that the model outperformed other feature sets. The authors used their self-generated dataset for this work. Lotfollahi et al. [81] compared the experimental performance of CNN and Stacked Auto-encoder (SAE) for encrypted traffic detection under the same dataset. 

\begin{figure*}
\centerline{\includegraphics[width=42pc]{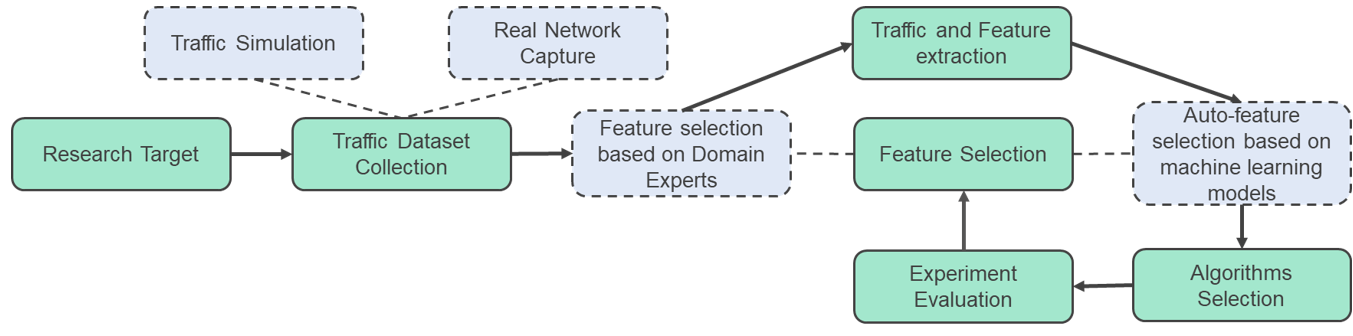}}
\caption{General framework of a machine learning based encrypted traffic detection model}
\end{figure*}

Khalife et al. [79] provided a fairly comprehensive framework on network traffic classification. They divided traffic detection and classification into three stages, namely data input, classification technique selection, and final result output. For each stage, they listed the current technologies and methods in a general way. In the classification technique selection stage, there are machine learning technology methods, statistical methods (i.e., Heuristics), and graphical technique methods (e.g., Motifs and Graphlets). However, the paper lacks an in-depth technical analysis of the machine learning based encrypted traffic classification techniques. In this work, we analyse the process of machine learning based encrypted traffic classification from different studies and formulate our framework for the process.

\section{\bf METHODS REVIEW BASED ON OUR FRAMEWORK}

Fig. 2 illustrates our proposed framework. It is specifically designed to suit a variety of detection models and is compatible with most existing studies, regardless of the datasets, features or algorithms used. The framework consists of 6 main steps. In the first step, the research targets should be set. This is followed by traffic dataset collection which can be done via two methods: real traffic collection and simulated traffic generation. The third step depends on the mode of feature selection employed. Feature selection may be performed manually through the advice of domain experts or in an automated manner using machine learning models. If the former is used, feature selection will be performed first and the traffic and features extraction is done next based on the selected features. On the other hand, if the latter is used, all possible features are first extracted from the traffic followed by automated feature selection to obtain the most ideal feature sets. After feature extraction and selection, the selection of algorithms is done as the next step. Finally, experimental evaluation is performed using the selected features and algorithms. We will discuss and review methods proposed in different research based on these steps in this section.

\subsection{\bf Research Objective}

Encrypted traffic classification can be generally divided into two subcategories: 1. encrypted \& unencrypted traffic classification. 2. encrypted malicious traffic detection.

The research on encrypted \& unencrypted traffic classification may also be regarded as the initial step of the research on encrypted malicious traffic detection. This is needed to successfully extract encrypted traffic from mixed traffic of encrypted and unencrypted traffic. Yao et al. [15] proposed an LSTM with attention mechanism RNN model for the identification of encrypted traffic, which can avoid complex and time-consuming feature engineering and learn the relationships among traffic flow features at the same time. Niu et al. [16]  proposed a heuristic approach which combines both statistics and machine learning to classify the encrypted network traffic, so as to make up for their respective shortcomings. The modified National Institute of Standard and Technology (NIST) test suite is utilized by the authors to simplify the interactions between packet processing and feature extraction. A handshake-skipping algorithm (HST-R) is also proposed to skip the handshake process during feature extraction, so as to avoid bringing the wrong message to statistical-based approaches. It also further increases the classification accuracy of the proposed machine learning based classifier. Zhao et al. [73] proposed an algorithm to perform the identification of encrypted traffic from both public and private encryption protocols which is named as encrypted traffic identification based on weighted cumulative sum test (EIWCT). The method utilizes the EIWCT to process each new incoming packet instead of testing after all packets arrived, as well as identifies the traffic based on weighted conflated results. The proposed algorithm reduces computation time and achieves the online identification of encrypted traffic from common encrypted protocols, such as SSL and SSH, and other encrypted private protocols efficiently. Above 90\% accuracy is achieved for SSL and private encryption protocol traffic in their experiments. Shahbar et al. [94] conducted comparative experiments based on supervised learning algorithms such as RF and Naïve Bayes for The Onion Router (Tor) and Invisible Internet Project (I2P) encrypted traffic classification.

For the research on encrypted malicious traffic detection, various approaches have been proposed in existing works. Many detection approaches are proposed regardless of the type of encrypted protocols. Stergiopoulos et al. [14] proposed a machine learning based encrypted malicious traffic detection model by using only a few features. The authors compared 7 different algorithms on their traffic datasets with multiple protocols to prove they use fewer features, fewer data and shorter time to achieve similar or even higher performance compared with other existing works. Furthermore, there are also many works focusing on certain encryption protocols. Most research aim to detect encrypted malicious traffic under HTTPs protocols, because HTTPs has become the most commonly used encryption mechanism for both legitimate and malicious traffic encryption. Research [11][13][25][28][29][33][92] extracted unique features from TLS/SSL protocol to train their HTTPs detection models. Shekhawat et al. [29] compared SVM, XGBoost, and RF algorithms with TLS/SSL features for binary detection and multi-classification of encrypted malicious traffic, and finally proposed that XGBoost with selected TLS/SSL features can achieve a 99.15\% accuracy under their own training dataset. Anderson and McGrew et al. [28] combined the unique features of TLS/SSL protocol and statistical numerical features of the encrypted traffic to detect HTTPs malicious traffic. Yang et al. [8] reassembled SSL records from packets and proposed an SSL malicious traffic detection method by deep learning combination of LSTM, AE, and CNN. Torroledo et al. [90] proposed a deep learning based identification system to detect legitimate and malicious TLS traffic certificates by utilizing the TLS certificate content. Meghdouri et al. [80] proposed an RF detection model for TLS and Internet Protocol Security (IPSec) protocols with novel cross-layer features, which comprise of features in the application level, conversation level and endpoint behaviour level at the same time. 

There are many works on the binary classification of encrypted traffic as malicious or legitimate [6]. For example, SSL malicious traffic detection model [8] classified SSL traffic as malicious or legitimate without considering the malware families. Side-channel features based encrypted malicious traffic detection model [14] focused on binary classification. HTTPs malicious traffic detection model in [28] is to classify traffic into enterprise and malware traffic. Some works also proposed multi-class classification of traffic belonging to different malware families. For example, Liu et al. [12] proposed a framework, which is comprised of detection models for 24 malware types. However, the performance of the models in the framework was not tested separately. If any model in the set of detection models performs badly, it is assumed that the other models with good performance will buffer the overall result. Zeng et al. [9] proposed a framework of encrypted traffic multi-class classification and intrusion detection based on deep learning. The framework is constructed from three different deep learning algorithms, CNN, LSTM, and SAE. However, there is no existing work that covers all types of traffic and malware families. The performance of multi-class classification is also observed to be generally worse than the binary classification approaches. 

In all, existing research can be summarised in four general approaches: 

1. Binary classification regardless of the encryption protocols. 

2. Binary classification under certain encryption protocols. 

3. Multi-class classification based on malware families regardless of the encryption protocols. 

4. Multi-class classification based on malware families under certain encryption protocols. 

Meanwhile, the encrypted traffic classification can also be applied to IoT devices domain. IoT devices are generally resource limited and more susceptible to cyber attacks. Due to the resource constraints and heterogeneity of IoT devices, traditional endpoint and network security solutions are not a good fit for securing IoT devices. Six challenges are summarised in securing IoT devices by [20]. They are endpoint security solutions with limited support, challenges in protecting device-to-device communications, high IoT devices diversity, high security deployment and operational costs, privacy, performance and attack risk trade off, and that most edge networks are vulnerable to attacks. Nakahara et al. [19] proposed an anomaly detection system that avoids relying on IoT device itself to detect and analyse packets, but rather by using other devices such as a low load home gateway. Then, the statistical features of aggregated traffic information are used, so as to reduce the processing load of the home gateway and the model. Wang et al. [89] proposed a deep learning based encrypted traffic classifiers, named as DataNets. MLP, SAE, and CNN are utilized in the classifier for encrypted traffic classification on software defined network. Aceto et al. [27] proposed a mobile traffic classifier which can deal with encrypted traffic through automatic feature extraction and deep learning.

Finally, in the process of setting our research objectives, we also need to consider the running environment of the detection model. That is, whether a detection model runs online or offline because different operating environments have different requirements for efficiency, robustness, and accuracy. At times, we need to compromise one requirement to better meet the others.

\subsection{\bf Traffic Dataset Collection}
After the research objectives have been established, the next step is to collect the corresponding datasets for model training. Data collection methods can be generally divided into two types: real traffic collection and simulated traffic generation. For real traffic collection, traffic data is collected from an operational network environment through commonly used traffic collection software, such as Wireshark. The collected data is then analysed and labelled for use. In [71], the authors deployed the IoT honeypot to collect the network traffic dataset from 2017 to 2018. In this one and a half years, the honeypot was presented on 40 public network IP addresses and the traffic was forwarded to 11 real IoT devices. In another project [58] a sample of internet traffic from Japan to USA over different time periods is collected. The data collection began in 2001. UNIBS dataset [65] contains 27GB traffic, which is captured on the university campus network for three consecutive working days. The real traffic collection method ensures the authenticity of the network traffic in the dataset, but it also has some shortcomings. Due to the low frequency of real network attacks compared to normal network interaction, the collected dataset is often imbalanced and has a lot of redundant data, which requires a lot of time for screening and labelling.

The simulated traffic generation usually generates targeted traffic datasets through the construction of a simulated network or the use of a script simulator. Yu et al. [74] built a network structure which includes a gateway server and several intranet clients. Metasploit is then used to generate different attacks to the gateway of servers, and Wireshark is utilized to collect traffic at the gateway of servers. In [53], part of traffic data is also generated by simulation. Datasets collected and generated by this method often contain more attacks and malware types. They are also more balanced than the datasets from real traffic collection, and are more suitable for machine learning. However, the datasets generated from simulated traffic generation may differ from real network data, and may not truly represent the real network traffic conditions.

Training an encrypted malicious traffic detection model requires a dataset to have following features: 

1. Sufficient amount of encrypted traffic.

2. High variety of encrypted malicious attacks.

3. Class balanced.

4. Ground truth is confirmed.

5. No redundant data.

Unfortunately, there is no well-established dataset for this specific domain problem currently. Thus, research works often use their own private datasets or select from the few available public datasets. This presents a challenge of performing fair performance comparisons among these existing works. Therefore, we aim to address this issue so as to help lay a good foundation to provide for future research on this topic.  We collect the publicly available datasets which are applicable to encrypted traffic analysis and analyse their targeted and applicable fields, characteristics, limitations, and whether they contain encrypted malicious traffic. We show the compiled dataset list in Table I. Through our analysis, we found that most public datasets either contain a few or do not contain encrypted traffic at all. Datasets containing encrypted malicious traffic are even rarer. 

\begin{table*}
\centering
\caption{The summary of public traffic datasets}
\begin{tabular}{|l|l|l|l|c|c|} 
\hline
Dataset Title                                                                           & \begin{tabular}[c]{@{}l@{}}Year of\\Release\end{tabular} & Characteristics and Limitations                                                                                                                                                                                                                                                                                                                                                                                                                                                                                                                                                                                                                                  & \begin{tabular}[c]{@{}c@{}}Contain \\Encrypted\\Malicious \\Traffic\end{tabular}  \\ 

\hline
\begin{tabular}[c]{@{}l@{}}IoT-23 \\ Dataset [51]\end{tabular}                          & 2020                                                                      & \begin{tabular}[c]{@{}l@{}}20 malwares traffic are captured in IoT devices. It also includes 3 captures of legitimate \\ IoT devices traffic. However, only a very small part of traffic is encrypted. The dataset \\contains only raw data. \end{tabular}                                                                                                                                                                                                                                                                                                                                                                                            & Y                                                                                \\ 
\hline
\begin{tabular}[c]{@{}l@{}}IoT \\ Encrypted \\ Traffic [50]\end{tabular}                & 2020                                                                      & \begin{tabular}[c]{@{}l@{}}The dataset contains traffic from several IoT devices. Most traffics are unencrypted, \\which are not applicable to encrypted traffic detection model training. The dataset \\contains only raw data. \end{tabular}                                                                                                                                                                                                                                                                                                                                                                                         & N                                                                                 \\ 
\hline
\begin{tabular}[c]{@{}l@{}}UNSW\_NS \\ 2019 [70]\end{tabular}                           & 2019                                                                      & \begin{tabular}[c]{@{}l@{}} The dataset contains encrypted legitimate and malicious traffics from more than 28 \\different IoT devices. Features extracted from captured raw data are not provided.\end{tabular}                                                                                                                                                                                                                                                                                                                                                               & Y                                                                                 \\

\hline
\begin{tabular}[c]{@{}l@{}}MIRAGE-2019 \\ dataset [31]\end{tabular}                           & 2019                                                                      & \begin{tabular}[c]{@{}l@{}} The dataset is collected by more than 280 experimenters in the ARCLAB laboratory \\at the University of Napoli "Federico II". It is a human-generated mobile-app dataset \\from May 2017 to May 2019 for mobile traffic analysis based on the authors proposed \\MIRAGE architecture[31]. The dataset contains 40 Android apps from 16 different \\categories.\end{tabular}                                                                                                                                                                                                                                                                                                                                                               & N                                                                                 \\

\hline
\begin{tabular}[c]{@{}l@{}}CIC-AndMal \\ 2017 [60]\end{tabular}                   & 2017                                                      & \begin{tabular}[c]{@{}l@{}}The dataset is released on Canadian Institute for Cybersecurity[18]. 5000 of the collected\\ samples (426 malware and 5065 legitimate) were installed on real\\devices. All samples are categorized into 4 malware types, from 42 malware families. \\Limited extracted features are provided using CICFlowMeter [78].\end{tabular}                                                                                                                                                                                                                                                                                                                                                                           & Y                                                                               \\ 
\hline
\begin{tabular}[c]{@{}l@{}}CICIDS \\ 2017 [53]\end{tabular}                       & 2017                                                      & \begin{tabular}[c]{@{}l@{}}legitimate and common attacks traffic, which resembles the real traffic (PCAPs) are \\captured in this dataset. legitimate traffic is generated by simulation. Limited features\\are extracted from captured raw traffic.\end{tabular}                                                                                                                                                                                                                                                                                                                                                                                                      & Y                                                                               \\ 
\hline
\begin{tabular}[c]{@{}l@{}}VPN-\\ non-VPN\\ trafﬁc dataset \\ {[}55]\end{tabular} & 2016                                                      & \begin{tabular}[c]{@{}l@{}}A regular traffic session and a session over VPN were captured. There are 14 traffic\\categories in this dataset. Captured data have been pre-processed and ISCXFlowMeter\\(old version of CICFLowMeter [78]) is utilized to create features csv files.\\ Selected features file and original PCAP file are provided.\end{tabular}                                                                                                                                                                                                                                                                                                                  & N                                                                               \\

\hline
\begin{tabular}[c]{@{}l@{}}UNSW\_NS \\ 2015 [69]\end{tabular}                           & 2015                                                                      & \begin{tabular}[c]{@{}l@{}}The available dataset is generated by the IXIA PerfectStorm tool [69]. It contains 49\\features of 6 feature categories and ten traffic classes (Normal, Fuzzers, Analysis, \\Backdoors, Denial-of-Service, Exploits, Generic, Reconnaissance, Shellcode, and Worms). \end{tabular}                                                                                                                                                                                                                                                                                                                                                        & Y                                                                                 \\

\hline
\begin{tabular}[c]{@{}l@{}}FIRST 2015 \\ {[}64]\end{tabular}                      & 2015                                                      & \begin{tabular}[c]{@{}l@{}}Trafﬁc is composed from Reverse Shell shellcode connections, website defacing \\ attacks, ransomware downloaded attack crypto locker and a command and conquer \\ exploit attack (C2) over SSL that takes over the victim machine. However, since \\encrypted malicious traffic in the dataset is significantly less than other forms \\of malicious traffic [14], it is not suitable for encrypted malicious traffic detection.\end{tabular}                                                                                                                                                                                                                                                                              & Y                                                                               \\ 

\hline
\begin{tabular}[c]{@{}l@{}}Malware \\ Capture \\ Facility \\ Project [49]\end{tabular}  & 2013                                                                      & \begin{tabular}[c]{@{}l@{}}The malware in this dataset is executed under bandwidth limit and spam interception.\\ The most important characteristic of this dataset is malware is executed in a long \\ periods of time (up to several months). The dataset contains only raw data.\end{tabular}                                                                                                                                                                                                                                                                                                                                                 & Y                                                                                 \\ 
\hline
\begin{tabular}[c]{@{}l@{}}Malware \\ traffic \\ analysis net\\ {[}52]\end{tabular}     & 2013                                                                      & \begin{tabular}[c]{@{}l@{}}Datasets are from malware traffic analysis website. The website provides lots of \\ different malware traffic since 2013. However, traffic data is complex and time \\consuming to process. The dataset contains only raw data.\end{tabular}                                                                                                                                                                                                                                                                                                                                                         & Y                                                                                 \\ 
\hline
\begin{tabular}[c]{@{}l@{}}CICIDS \\ 2012 [54]\end{tabular}                       & 2012                                                      & \begin{tabular}[c]{@{}l@{}}A systematic approach to generate HTTP, SMTP, SSH, IMAP, POP3, and FTP traffic \\datasets is used. Thus, the datasets cannot fully represent real network traffic. \\Limited features are extracted from captured raw traffic.\end{tabular}                                                                                                                                                                                                                                                                                                                                                                                                          & Y                                                                               \\
\hline
CTU-13 [48]                                                                             & 2011                                                                      & \begin{tabular}[c]{@{}l@{}}This dataset contains a total of 13 botnet capture in different scenarios. Each of these\\ captures are collected for a period of time. Most traffics are unencrypted traffic, but \\ contain encrypted traffic. The dataset contains only raw data.\end{tabular}                                                                                                                                                                                                                                                                                                                                                     & Y                                                                                \\ 

\hline
\begin{tabular}[c]{@{}l@{}}Webldent \\ 2 Traces \\ {[}57]\end{tabular}                  & 2006                                                                      & \begin{tabular}[c]{@{}l@{}}Web requests and responses over an encrypted SSH tunnel were collected. Limited \\features are extracted from captured raw traffic.\end{tabular}                                                                                                                                                                                                                                                                                                                                                                                                                                                                                      & Y                                                                                 \\ 

\hline
\begin{tabular}[c]{@{}l@{}}Kyoto \\ Dataset \\ {[}63]\end{tabular}                      & 2006                                                                      & \begin{tabular}[c]{@{}l@{}}The dataset only contains the traffic on honeypots, and the captured raw traffic PCAP file \\is not provided. It contains limited extracted features csv files only, which means we \\cannot further process features from captured raw traffic.\end{tabular}                                                                                                                                                                                                                                                                                                                                                                                                          & Y                                                                                 \\ 

\hline
\end{tabular}
\end{table*}

\subsection{\bf Traffic and Feature Extraction}

\paragraph{\bf \textit{Traffic data pre-processing}}

The first step of dataset pre-processing is data cleaning and filtering. Most public datasets are in the format of Packet Capture (PCAP)/ PCAP Next Generation (PCAPNG), and these PCAP/PCAPNG files often record the original raw traffic data. For this kind of data, it is necessary to clean up irrelevant network packets which are not applicable in the research of encrypted malicious traffic detection, such as Address Resolution Protocol (ARP) or Internet Control Message Protocol (ICMP) packets. Then we need to remove duplicated, damaged, unnecessary, and incompletely captured traffic streams or information which may interfere with model training. Data truncation and padding are utilized to keep the length of input data consistent [83].  Finally, we filter out the non-encrypted traffic which results in the final dataset containing only encrypted traffic. There are also some public datasets that are not released under basic PCAP files, but csv format with prior cleaning, filtering, and feature extraction, such as [53]. However, if there is no label in csv feature datasets to indicate if these features are extracted from encrypted traffic or unencrypted traffic, and the source of the dataset does not provide the original traffic files (i.e., PCAP files), such datasets are difficult to the encrypted traffic detection model training. Furthermore, if a feature needed to train the model is not included in this kind of dataset, such as in [63], the dataset cannot be used as well. In addition, we also need to convert or encode categorical data, such as network protocol information, encryption protocol certificates, server indication names, and so on, in the dataset to numeric data.

Data imbalance in datasets is also a problem. In existing research, two methods are usually adopted to solve this problem. The first one is the combination of different datasets, or the use of self-generated dataset to increase the type and quantity of traffic. In the second method, data augmentation, over-sampling, under-sampling or over-sampling followed by under-sampling can be applied to address data imbalance. Yan et al. [26] utilized a mean synthetic minority over-sampling (SMOTE) to balance their own dataset. Wang et al. [89] utilized the under-sampling method to reduce the number of samples of major classes in the dataset. Vu et al. [91] conducted a series of comparative experiments with different existing methods of addressing data imbalance. Data normalization is another important part of data pre-processing, especially for deep learning. Data normalization can normalize the traffic data or feature value to a range of zero to one or negative one to positive one so as to reduce data redundancy, enhancing the integrity and efficiency of model training.

\paragraph{\bf \textit{Feature extraction}}

Feature extraction is the next key step after dataset collection. In order to select the best feature set, it is necessary to first understand the types and characteristics of features that can be extracted from traffic data. At present, there is no recognized classification or naming convention for network traffic features. Thus, we performed analysis and categorization of features by studying the relevant works. The study led us to broadly classify the features into two types: protocol-agnostic numerical features and protocol-specific features.

Protocol-agnostic numerical features have two granularities, which are packet based features and session based features. Packet based features are features extracted at the traffic packet level, such as time difference between packets in a session flow, packet size (or packet length) of each packet, payload size (or payload length) of each packet, and value changes of TCP windows length per session. Session based features are features extracted at the session flow level, such as session flow duration, total bytes from client/server in each session, total number of packets from client/server in each session, and total length of IP packet header per session. 

Feature engineering can be further applied to these packet based and session based features. Statistical features are obtained from conducting statistical calculations on such features. Common calculations include mean, median, maximum, minimum, variance, and standard deviation (STD) of packet based and session based features. [19][14] selected packet based features such as packet size, payload size and the time difference between sessions as their features to train their detection models. In [20][32][76][77], the authors selected session based features and statistical features as their feature set. Liu et al. [12] extracted more than 80 packet based, session based, and statistical features, which are used in their detection models.

Many works selected their unique subsets from protocol-agnostic numerical features as their model training feature set based on the distinctive attributes of features. Celik et al. [17] defined tamper resistant features (e.g., IPratio and goodput), such features are not related to port, flag or payload. These features are consistent and difficult to be masked by adversaries. Stergiopoulos et al. [14] defined their TCP side channel features (e.g., ratio to previous packets), which reduces the size of training dataset and training time while ensuring encrypted malicious traffic detection accuracy is above 99\%. Liu et al. [12] divided their extracted features into 4 different sub-categories based on the attributes of features: TCP/IP header features (e.g., IP header); time based features (e.g., average inter-arrival time of packets), length related features (e.g., packet length and payload length), and packet variation features (e.g., TCP window change times). These features are used for clustering. Then, different feature subsets are selected based on sequential forward selection (SFS) algorithms for their more than 20 supervised learning models. Sarkar et al. [77] selected time related features, such as flow duration, mean of backward packet time difference, STD of forward packet time difference, and STD of time difference for their model. 

For protocol-specific features, features from the information of encrypted protocols are extracted instead. Such features frequently appear in the research of HTTPs traffic detection as HTTPs is dominating as the encrypted traffic type currently. Protocol-specific features extracted from HTTPs protocols are mainly referred to as TLS/SSL features. Such features are usually extracted from three different log files, conn.log file, SSL.log file, and X509.log (certificate log) file. One of the most commonly used log generation applications from captured raw network traffic is Zeek IDS (Bro IDS). Zeek IDS generates such log files based on traffic and protocols by processing raw network traffic [24]. These log files are interconnected to one another through the uid (unique ids) column and cert\_chain\_fuid (certificate identifiers) column. Features extracted from conn.log file have certain overlaps with the protocol-agnostic numerical features. For example, its features also include the number of flow and payload bytes from client and server. Ssl.log file and x509.log file are interconnected. Extracting x509 features is reliant on ssl.log information, and the relationship among these three files is explained in Fig. 3. Examples of ssl Log Features include TLS version types and the ratio of the same issuer. X509 Log Features include public certificate key mean, mean of certificate validity, and average number of domains in Subject Alternative Name (SAN). [11][13][25][28][29][33][92] used features from TLS/SSL features in their works. In [11], 12 connection features, 6 SSL features, and 6 x509 (certificate) features were selected as the TLS/SSL feature set for CNN and RF models. In [28][33], the authors chose to combine protocol-agnostic numerical features and TLS/SSL features as their feature set. 

\begin{figure}
\centerline{\includegraphics[width=21pc]{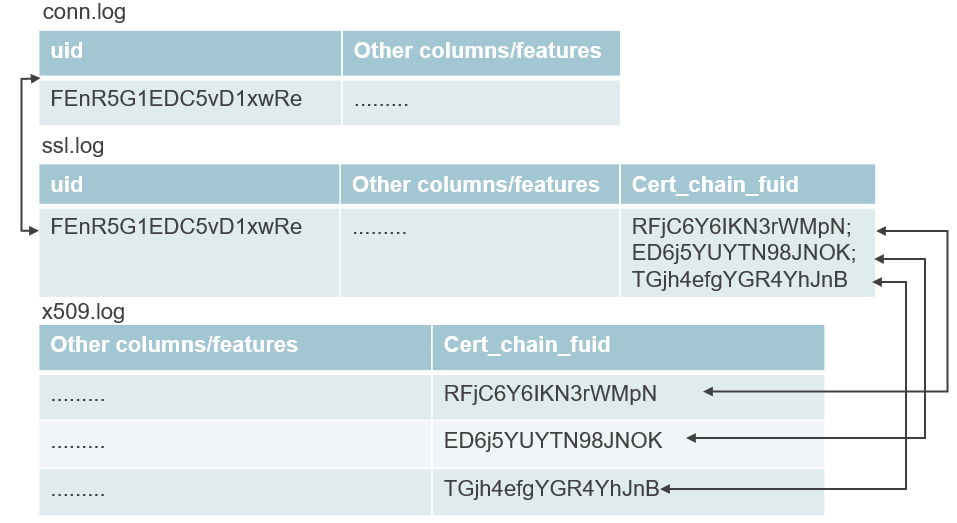}}
\caption{Relationship among protocol logs}
\end{figure}

For protocol-specific features, one significant limitation is that such features are only applicable to their related encrypted protocols. Therefore, if the dataset contains other encryption protocols, protocol-specific features will not be applicable anymore. On the other hand, the extraction of protocol-agnostic numerical features does not depend on any specific contents or information of the traffic communication and protocols. Thus, no matter what encrypted mechanism is utilized, protocol-agnostic numerical features can always be extracted. However, since there is a large amount of protocol-agnostic numerical features with different extraction logic, the extraction process is complex, time-consuming, and requires prior knowledge.

\subsection{\bf Feature Set Selection}

It is important to note that more features do not imply better results. Having too many features in the feature set requires a high volume of memory and computation, extends the model training time and may also affect the model accuracy. 

Common feature selection methods can be divided into two categories. The first one is domain expert based selection. Domain experts select a set of features that they think is the most appropriate based on their experience and knowledge. After that, such suggested features will be extracted directly from datasets and used as model input. In [28], the authors found that iterating the initial feature set and adding domain experts’ suggested features can greatly enhance the detection system’s performance. The RF ensemble method outperformed competing methods from an algorithmic perspective based on a feature set from the domain experts. They also mentioned that as compared with algorithm selection, feature engineering is more decisive and important. However, because of privacy concerns, the authors did not disclose their datasets or domain experts’ suggested feature set. Many research [14][16][17][19][28][29][33] have used similar ways to artificially select features for their experiments. However, it is not always reliable to manually choose the most important features. Human error may exist and human may not be able to find non-intuitive features, which may greatly affect the performance of the detection models. Noted that for domain expert based selection, the features will first be selected and this set of features will then be extracted from the traffic.

Another category of feature selection methods is based on machine learning self-selection with some algorithms. It requires humans to extracted all possible features and selected algorithms will rank and select the most suitable features as the final feature set, or directly process raw data to learn and extract required features by themselves. In [11], after the image augmentation and feature extraction, the authors utilized the mean decrease in impurity algorithm to rank features based on their importance. Lastly, they generated a feature set with top ranked features. Liu et al. [12] applied SFS algorithm to increase the size of feature set gradually until they found their optimal feature set. However, SFS uses a greedy algorithm, which may fall into the local optimum and is very time consuming. To avoid these limitations, the authors also further enhanced the SFS algorithm by adding random selection in each round. Finally, they combined feature sets with top performance to get their final feature set. Shekhawat et al. [29] extracted TLS/SSL features and applied recursive feature elimination (RFE) to eliminate features with the lowest ranking score iteratively. The author applied the final feature set to XGBoost, SVM, and RF algorithms. For RF model, they obtained a nearly 99\% accuracy while XGBoost achieved 99.15\% accuracy. Desai et al. [30] proposed a feature ranking framework based on fundamental statistical tests. It is specifically designed for IoT device traffic classification, which can select important features from IoT network traffic, reducing cost and protecting user privacy. Their experiments indicate that the small number of features can achieve similar accuracy as compared to other existing methods. Many research [13][20][32] are designed to use machine selection on the optimal features chosen from a large set of extracted features without human intervention. On the other hand, [9][10][37][44][47] directly used raw data as the deep learning methods input. The model can learn and extract the required features automatically. Zheng et al. [9] applied CNN, LSTM, and SAE to learn features from different aspects. For example, extracting features from time based aspect was based on LSTM in [9]. Machine learning based feature selection can discover non-intuitive features, and can avoid human errors and bias at the same time. For this method, manual feature extraction process is skipped and raw data is fed to automated feature selection and extraction, or feature extraction is performed first to provide a full set of features which is then fed to automated feature selection. However, due to the black box feature of AI, especially when deep learning algorithms are involved, it is very challenging to interpret the feature selection process and provide explainability of the reasons features were selected as optimal. Additional research is required to solve the above issue [62][67]. 

\subsection{\bf Algorithms Selection}

Many mature traditional methods for traffic detection, such as DPI, are no longer applicable to encrypted traffic. At present, traditional machine learning methods and deep learning methods are two mainstream research directions in this area.

\paragraph{\bf \textit{Traditional Machine Learning Approach}}

For the traditional machine learning, algorithms and feature set optimization are the main focus. For unsupervised learning, there are many works like Chen et al. [13], which proposed a three-stage hierarchical sampling approach by further developing density peaks clustering algorithm (THS-IDPC) based on grid screening, custom centre decision value and mutual neighbour degree (DPC-GS-MND). Experiments have shown that their DPC-GS-MND is better than average-linkage [86], DPC, modified density peak clustering algorithm (MDPCA) [87] and DPC-GS. Furthermore, their performance of THS-IDPC based model is shown to be better than selective sampling [84], smart sampling [85] and hierarchical clustering based sampling (HCBS) [86]. The algorithm can significantly reduce the cost of computation and enhance the accuracy and efficiency of encrypted malicious traffic detection model. However, the method in this work cannot handle the class imbalanced problem. Furthermore, DPC-GS-MND uses the K nearest neighbour idea, but the K value has to be decided manually. 

Hafeez et al. [20] proposed an IoT Keeper model which is constructed by the fuzzy C-means (FCM) clustering algorithm with the fuzzy interpolation scheme, to perform malicious IoT traffic classification. Such unsupervised learning algorithm does not need fully labelled dataset as the model training input. Furthermore, the proposed model will not be restricted to certain types of IoT devices. A novel mechanism, Adhoc Overlay networks, is also applied to the proposed model, which can strengthen the access control to IoT devices network activities actively. However, IoT Keeper was not tested with encrypted malicious traffic and the authors listed that as a future research direction. Unsupervised deep learning like AE is further discussed in the next deep learning approach subsection.

On supervised learning, Meghdouri et al. [80] proposed an RF classification model based on a multi-key based approach, which is a novel cross-layer feature representation of traffic data under TLS and IPSec protocols. They tested the model using three different datasets, CICIDS-2017[54], UNSW0-NB15[22] and ISCX-bot-2014[72], which achieved 100\%, 92.6\% and 99.2\% F1 scores, respectively. Comparative experiments were also conducted with other existing methods which used the same datasets. They showed that their model outperformed those methods. 

Stergiopoulos et al. [14] defined their side channel features and applied their feature set into 7 different machine learning algorithms to verify the performance of these side channel features. The paper focuses on malicious traffic detection, but also includes encrypted traffic detection. The encrypted dataset was extracted from CTU-13 [48], FIRST [64], and Milicenso [66] and they obtained 99.8\% accuracy under CART model. Shekhawat et al. [29] tested XGBoost, SVM and RF algorithms with a machine selected feature set and achieved 99.8\% accuracy with the XGBoost model. 

Ma et al. [32] proposed an enhanced KNN algorithm, which is named WKNN-Selfada (feature weight self-adaptive algorithm for weighted feature KNN). The proposed algorithm improved the KNN distance calculation and also included a sub-algorithm that can choose the suitable feature set and feature weights. However, although the paper mentioned that the proposed algorithm can be used for encrypted traffic detection, they did not perform any experimental evaluation.

Niu et al. [16] proposed a heuristic statistical testing (HST) approach. The HST approach consists of 3 parts: Jnetpcap-based Executor (handshake-skipping algorithm), enhanced NIST Test Suite and an algorithm classifier (such as C4.5). The experiment data used in this paper were based on two proprietary protocols (Freegrate and Ultrasuff) and one private custom unknown protocol. The dataset was not publicly released. A 10-fold cross validation was used in the experiment. The experiment shows that after applying the handshake skipping algorithm, the accuracy of the experiments improved significantly and that HST is more robust than the other tested machine learning methods. 

Frameworks that combine supervised learning and unsupervised learning also exist in many works. In [12] the authors proposed a distance based encrypted malicious traffic identification framework, which comprised of a series of detection models for different malware types. They firstly applied the Gaussian mixture model (GMM) and ordering points to identify the clustering structure (OPTICS) to classify the new malware class (FClass) based on the distance calculation between malwares. After that, 24 XGBoost encrypted malicious detection models are trained to identify the 24 kinds of malware. Comparative experiments have shown that their proposed method is better than the MalClassifier model in [96] and the CluClas model in [21].

\paragraph{\bf \textit{Deep Learning Approach}}

Research on deep learning has grown rapidly in recent years and has achieved remarkable results. Deep learning based encrypted traffic detection has many obvious advantages, such as the ability to automatically extract the required data features through its own feature learning. It is also easier to find non-intuitive connections among traffic features that humans cannot.

A general deep learning based framework is proposed by Aceto et al. [98] for encrypted and mobile traffic classification. The proposed framework provides clear guidelines for designers in deep learning based traffic classification. It also overcomes the design limitations of the existing single-modality or single-task learning methods by jointly using multimodal and multi-task techniques. Moreover, the authors test two implementations of the framework based on three datasets, which are generated by the activity of mobile users. Aceto et al.[99] also proposed a novel multimodal and multitask deep learning based approach for multipurpose encrypted traffic classification, DISTILLER. The proposed method overcomes the performance limitations of state-of-the-art single-mode deep learning models based on heterogeneous and structured traffic. It can also tackle different traffic categorization problems from different providers. A fair comparative experiment with 8 other existing deep learning based encrypted traffic classification methods is conducted by using the VPN-non-VPN public dataset[55]. Experiment results indicate that DISTILLER has a better performance than other methods.

Bazuhair et al. [11] proposed a new method which can enhance the generalization of CNN in encrypted malicious traffic detection. That work focused on HTTPs malicious traffic and the authors developed a binary detection model. A new encoding method which can convert TLS/SSL features into images was proposed. Then, Perlin noise is utilized to do the data argumentation. This enhanced the generalization of the deep learning model. CTU-13 [48] dataset is used in the experiment and the CNN model achieved 97\% accuracy, with a 0.4\% false negative rate and 5.6\% false positive rate (FPR). Lucia and Cotton. [46] proposed a TLS malicious traffic classification based on a public dataset malware capture facility project[49]. SVM and CNN were selected to conduct the experiment. The one-dimensional CNN achieved 99.91\% for both accuracy and F1 with Adam optimizer, 32 batch size, 100 epochs and early stopping. The non-linear SVM with radial basis function kernel achieved 99.97\% for both accuracy and F1. That work was an enhancement of their previous work [82]. Wang et al. [37] proposed an end to end encrypted traffic classification method. VPN-non-VPN [55] traffic data was utilized. The data was split into the same length and used as the model input. Furthermore, the end-to-end framework merged feature extraction, selection and classifier processes, so as to automatically learn and discover the required features for classification. The experiments showed that the one-dimensional CNN outperforms the two-dimensional CNN. [34][43-45][88] also applied CNN in their experiments which achieved above 93\% detection rates.

Yao et al. [15] proposed two encrypted traffic classification methods. The first method is based on LSTM with attention mechanism and the second one is based on the hierarchical attention network (HAN). VPN-non-vpn [55] dataset was used for their comparative experiments with attention based LSTM, HAN, Deep Packet [81], one-dimensional CNN model [37], decision tree [35] and XGBoost. The experiment result indicated that their attention based LSTM model and HAN neural network model both outperformed the machine learning based model from [35] and the one-dimensional CNN model from [37]. Pascanu et al. [47] proposed a series of hybrid models that combined an Echo state networks (ESN) with a classifier (logistic regression or MLP) or RNN with a classifier (logistic regression or MLP). ESN plus logistic regression with maximum pooling for non-linear sampling outperformed other models. Prasse et al. [25] proposed an encrypted malware detection model based on LSTM. The work focused on HTTPs traffic and self-collected dataset by using cloud web security (CWS) and VirusTotal, which helped the authors get enough malicious and legitimate traffic. The proposed detection model can classify different malware families, even for new unknown malware. A comparative experiment was also conducted and showed that LSTM based model outperformed RF detection model.

There are also works that try to combine CNN and RNN to their models. Lopez-Martin et al. [83] combined RNN with CNN to classify IoT traffic regardless of the traffic was encrypted or not. The proposed model achieved a 95.74\% F1 score and 96.32\% accuracy in an imbalanced dataset. The research conducted model training using 5 different feature sets to analyse the importance of feature set selection in model training. The authors also considered the impact of the length of traffic session flow and the trade off between computing time and detection rate. Their comparative experiment using the different number of packets in each session indicated that keeping to between five and fifteen packets in each session flow can achieve above 94.5\% performance metric (accuracy, F1, precision, and recall). Furthermore, sessions with less than the pre-decided packets number will be padded as zeros to ensure each session has the same number of packets. Wang et al. [36] proposed a hierarchical spatial-temporal features-based intrusion detection system (HAST-IDS) by combining CNN with LSTM. Spatial features and temporal features can be learned by CNN and LSTM, respectively. The resultant model brought a reduction to the FPR. 

[7-10][22][38-42][75][81] applied AE in their detection models with their different self-collected or public datasets. A network intrusion detection(for both known and unknown attacks) based on a two-stage architecture is proposed by Bovenzi et al. [75], which is named H2ID. The first stage of this framework is to utilize a novel multimodal deep auto-encoder (M2-DAE) to perform a lightweight anomaly detection. The anomalous traffic is then classified into different types of attack traffic such as scans and distributed denial-of-service using soft-output classifiers in the second stage. BotIoT[23] dataset is selected to validate the performance of the proposed approach. In [81], comparative experiments based on CNN and SAE were conducted by using the VPN-non-VPN dataset[55]. CNN and SAE both achieved above 92\% accuracy for application and traffic classification. Yang et al. [8] proposed an encrypted malicious traffic classifier that combines CNN and LSTM auto-encoder. They finally achieved a 95.8\% detection rate. Zeng et al. [9] proposed an encrypted traffic based intrusion detection framework, named Deep-Full-Range (DFR). The framework is constructed by CNN, LSTM, and SAE to self-learn and extract features from raw data input. The authors conducted comparative experiments with KNN and other existing works which used the VPN non-VPN [55] and CICIDS 2012 [54] datasets to demonstrate that the proposed DFR is more accurate and robust. Xing et al.[10] proposed an online detection model based on deep dictionary learning, D2LAD, to address the noisy data label, long training time and high traffic data distribution variance. The model can learn and extract sequential features from raw traffic data input based on a pre-trained LSTM auto-encoder. The authors showed that their proposed work achieved a 94.5\% accuracy, which outperformed existing methods for online encrypted traffic detection. 

The research of applying deep learning provides us with many very useful detection methods. Such methods include the algorithms based on feature self-learning which were proposed to overcome the difficulty of traffic feature extraction and selection. These algorithms do not require human effort to extract traffic features. The algorithms then automatically extract the required features. On the other hand, research on applying deep learning to encrypted malicious traffic detection and classification are limited.

\section{\bf COMPARATIVE EXPERIMENT}

In this section, we perform two experiments to evaluate our experiment objectives. Specifically, our first experiment (referred to as Experiment 1) is to conduct a series of comparative experiments with different algorithms and feature sets to find a more reliable, consistent and fair result for each proposed existing work. By screening and analysing public datasets in Table I of Section IV.B.a., we curated a dataset composed entirely from public datasets that are applicable for encrypted malicious traffic detection. To the best of our knowledge, the dataset is more comprehensive and objective for use than those used by existing works. To prove this, a cross dataset validation is conducted as well in experiment 2. Moreover, this composed dataset also allows a fair comparison among the proposed works for encrypted malicious traffic detection and classification. The second experiment (or Experiment 2) aims to provide some insights on whether protocol-agnostic numerical features or protocol-specific features are better in performance, and how well one fares against the other.

\subsection{\bf Data Collection}

Existing research that adopts different public datasets may result in bias and the inability to compare results of proposed works in an objective manner. In order to aid in this effort to ensure that experiments can be conducted as fairly and comprehensive as possible, we aim to collect, analyse and derive data from publicly available datasets on the Internet, to compose a dataset for encrypted malicious traffic detection and classification. At the same time, we attempt to include datasets from as many different sources as possible to expand the variety of encrypted traffic.

Our dataset is composed based on three criteria: The first criterion is to combine widely considered public datasets which contain both encrypted malicious and legitimate traffic in existing works, such as Malwares Capture Facility Project dataset and CICIDS-2017 dataset. The second criterion is to ensure the data balance, i.e., the balance of malicious and legitimate network traffic and similar size of network traffic contributed by each individual dataset. Thus, approximate numbers of malicious and legitimate traffic from each selected public dataset are extracted. We also ensured that there will be no traffic size from one selected public dataset that is much larger than other selected public datasets. The third criterion is that our dataset includes both conventional devices' and IoT devices' encrypted malicious and legitimate traffic, as these devices are increasingly being deployed and are working in the same environments such as offices, homes, and other smart city settings.

Based on the criteria, 5 public datasets are selected from Section III.B.a. Table I. After data pre-processing, details of each selected public dataset and the final composed dataset are shown in Table II. Table II summarised the malicious and legitimate traffic size we selected from each selected public dataset by using random sampling, proportions of selected traffic size from each selected public dataset with respect to the total traffic size of the composed dataset(\% w.r.t the composed dataset), proportions of selected encrypted traffic size from each selected public dataset (\% of selected public dataset), and total traffic size of the composed dataset. From the table, we are able to observe that each public dataset equally contributes to approximately 20\% of the composed dataset, except for CICDS-2012 (due to its limited number of encrypted malicious traffic). This achieves a balance across individual datasets and minimizes bias towards traffic belonging to any dataset during learning. We can also observe that the size of malicious and legitimate traffic are almost the same, thus achieving class balance. To facilitate the research in this domain, we released our dataset[100] in Mendeley Data.

\begin{table*}
\centering
\caption{The selected public datasets in experiments}
\begin{tabular}{|l|l|l|l|l|l|l|l|} 
\hline
\textbf{Public Dataset}                                                                     & \begin{tabular}[c]{@{}l@{}}\textbf{\textbf{Year of}}\\\textbf{\textbf{Release}}\end{tabular} & \textbf{Type of Traffic}                                                & \begin{tabular}[c]{@{}l@{}}\textbf{Malicious}\\\textbf{Traffic}\\\textbf{Size}\end{tabular} & \begin{tabular}[c]{@{}l@{}}\textbf{legitimate }\\\textbf{Traffic}\\\textbf{Size}\end{tabular} & \begin{tabular}[c]{@{}l@{}}\textbf{Total }\\\textbf{Traffic}\\\textbf{Size}\end{tabular} & \begin{tabular}[c]{@{}l@{}}\textbf{\% w.r.t }\\\textbf{composed }\\\textbf{dataset}\end{tabular} & \begin{tabular}[c]{@{}l@{}}\textbf{\% of~}\\\textbf{selected}\\\textbf{public}\\\textbf{dataset}\end{tabular}  \\ 
\hline

\begin{tabular}[c]{@{}l@{}}UNSW NS \\ 2019 Dataset \\ {[}70]\end{tabular}                   & 2019                                                                                         & \begin{tabular}[c]{@{}l@{}}IoT Encrypted \\Traffic\end{tabular}         & \begin{tabular}[c]{@{}l@{}}12900 sessions\\193500 packets\end{tabular}                      & \begin{tabular}[c]{@{}l@{}}13300 sessions\\199500 packets\end{tabular}                        & \begin{tabular}[c]{@{}l@{}}26200 sessions\\393000 packets\end{tabular}                   & \textasciitilde{}22\%                                                                            & \textasciitilde{}60\%                                                                                      \\ 
\hline
\begin{tabular}[c]{@{}l@{}}CICIDS-2017\\ {[}53]\end{tabular}                                & 2018                                                                                         & \begin{tabular}[c]{@{}l@{}}Conventional\\Encrypted Traffic\end{tabular} & \begin{tabular}[c]{@{}l@{}}13000 sessions\\195000 packets\end{tabular}                      & \begin{tabular}[c]{@{}l@{}}13500 sessions\\202500 packets\end{tabular}                        & \begin{tabular}[c]{@{}l@{}}26500 sessions\\397500 packets\end{tabular}                   & \textasciitilde{}23\%                                                                            & \textasciitilde{}70\%                                                                                      \\ 
\hline
\begin{tabular}[c]{@{}l@{}}CIC-AndMal \\ 2017[60]\end{tabular}                              & 2018                                                                                         & \begin{tabular}[c]{@{}l@{}}Conventional\\Encrypted Traffic\end{tabular} & \begin{tabular}[c]{@{}l@{}}12403 sessions\\132859 packets\end{tabular}                      & \begin{tabular}[c]{@{}l@{}}12400 sessions\\186000 packets\end{tabular}                        & \begin{tabular}[c]{@{}l@{}}24803 sessions\\318859 packets\end{tabular}                   & \textasciitilde{}21\%                                                                            & \textasciitilde{}60\%                                                                                      \\ 
\hline
\begin{tabular}[c]{@{}l@{}}Malware Capture \\ Facility Project \\ Dataset [49]\end{tabular} & 2013                                                                                         & \begin{tabular}[c]{@{}l@{}}Conventional\\Encrypted Traffic\end{tabular} & \begin{tabular}[c]{@{}l@{}}13600 sessions\\204000 packets\end{tabular}                      & \begin{tabular}[c]{@{}l@{}}12180 sessions\\182700 packets\end{tabular}                        & \begin{tabular}[c]{@{}l@{}}25780 sessions\\386700 packets\end{tabular}                   & \textasciitilde{}22\%                                                                            & \textasciitilde{}50\%                                                                                      \\ 
\hline
\begin{tabular}[c]{@{}l@{}}CICIDS-2012\\ {[}54]\end{tabular}                                & 2012                                                                                         & \begin{tabular}[c]{@{}l@{}}Conventional\\Encrypted Traffic\end{tabular} & \begin{tabular}[c]{@{}l@{}}7613 sessions\\69648 packets\end{tabular}                        & \begin{tabular}[c]{@{}l@{}}6731 sessions\\71310 packets\end{tabular}                          & \begin{tabular}[c]{@{}l@{}}14344 sessions\\140958 packets\end{tabular}                   & \textasciitilde{}12\%                                                                            & \textasciitilde{}100\%                                                                                     \\ 
\hline
\textbf{Summary}                                                                            &                                                                                              &                                                                         & \begin{tabular}[c]{@{}l@{}}59516 sessions\\795007 packets\end{tabular}                      & \begin{tabular}[c]{@{}l@{}}58111 sessions\\842010 packets\end{tabular}                        & \begin{tabular}[c]{@{}l@{}}117627 sessions\\1637017 packets\end{tabular}                 &                                                                                                  &                                                                                                            \\
\hline
\end{tabular}
\end{table*}

\subsection{\bf Feature extraction and selection}

Protocol-agnostic numerical features and TLS/SSL features are extracted for experiments. Since there is no recognized optimal feature set, researchers have adopted different methods to select the feature set they think is the best, which has been discussed in Section IV. Therefore, for protocol-agnostic numerical features extraction, in order to compare the performance of models with different feature sets reliably, we extracted applicable features mentioned in research papers for encrypted traffic detection. A total of more than 113 unique protocol-agnostic numerical features were obtained. Table III shows a few commonly used features of protocol-agnostic numerical features extracted.

\begin{table}
\centering
\caption{Example of statistical numerical features}
\begin{tabular}{|l|} 
\hline
\textbf{Statistical numerical features}                      \\ 
\hline
Length of TCP payload                                        \\ 
\hline
Length of IP packets header                                  \\ 
\hline
TCP windows size value                                       \\ 
\hline
Time difference between packets per session                  \\ 
\hline
Interval of arrival time of forward traffic                  \\ 
\hline
Ratio to previous packets in each session                    \\ 
\hline
Total bytes from client in each session                      \\ 
\hline
Flow Duration of each session                                \\ 
\hline
Length of IP packets (Minimum; Maximum; Median; Mean; STD)   \\ 
\hline
Length of TCP payload (Minimum; Maximum; Median; Mean; STD)  \\ 
\hline
.........                                                    \\
\hline
\end{tabular}
\end{table}

As different works selected different feature sets, we conducted statistical analysis on the protocol-agnostic numerical features that appeared in those research and listed the features that appeared at high frequencies. In comparative experiments, the listed features will be constructed as a feature set, named Further Optimized Statistical (FOS) feature set.

5 different feature sets belonging to protocol-agnostic numerical features are used in experiments:

1. FOS feature set contains 14 features; 

2. Top 10 ranked features feature set from [12], which are the top 10 ranked features used in their detection framework by using enhanced SFS algorithm; 

3. Side channel feature set from [14], which contains five packet based features, but was shown in the existing work to have achieved a 99.8\% accuracy in their experiment; In our experiment, in order to use this packet based feature set to compare with other session based feature sets, we choose to only use the first 15 packets in each session. The reason for choosing 15 packets per session is based on the analysis result from [83] which states keeping between 5 and 15 packets in each session can achieve a better performance metric (also discussed in Section IV, E). This ensures the dataset contains as much data variety as possible and is suitable for both packet based feature set and session based feature set at the same time.

4. Tamper resistant feature set based on the idea from [17], which the selected features do not rely on TCP payload information or do not contain port and flag.

5. Time based feature set based on the idea from [77], where these time series relevant features appear at high frequencies in research.

Features in each selected feature set are shown in Table IV.

\begin{table*}
\centering
\caption{Features in each selected feature set}
\begin{tabular}{|l|l|l|l|l|l|} 
\hline
\textbf{Feature Name}                                                 & \textbf{1} & \textbf{2} & \textbf{3} & \textbf{4} & \textbf{5}  \\ 
\hline
mean TCP windows size value                                       & \checkmark          & \checkmark          &            &            &             \\ 
\hline
source port                                                           & \checkmark          & \checkmark          &            &            & \checkmark           \\ 
\hline
mean length of IP packet header                                       &            & \checkmark          &            &            &             \\ 
\hline
maximum interval of arrival time of forward traffic                   & \checkmark          & \checkmark          &            &            & \checkmark           \\ 
\hline
mean Length of backward IP packet header                               &            & \checkmark          &            &            &             \\ 
\hline
maximum interval of arrival time of backward traffic                  & \checkmark          & \checkmark          &            &            & \checkmark           \\ 
\hline
STD of backward packet length                          &            & \checkmark          &            &            &             \\ 
\hline
flow duration                                                         & \checkmark          & \checkmark          &            & \checkmark          & \checkmark           \\ 
\hline
time duration of backward traffic                                     &            & \checkmark          &            &            &             \\ 
\hline
total payload per session                                             &            & \checkmark          &            &            &             \\ 
\hline
destination Port                                                      &            &            &            &            & \checkmark           \\ 
\hline
STD of time difference between packets per session     &            &            &            &            & \checkmark           \\ 
\hline
minimum of time difference between packets per session                &            &            &            &            & \checkmark           \\ 
\hline
STD of interval of arrival time of backward traffic    & \checkmark          &            &            &            & \checkmark           \\ 
\hline
STD of interval of arrival time of forward traffic     &            &            &            &            & \checkmark           \\ 
\hline
minimum of interval of arrival time of backward traffic               &            &            &            &            & \checkmark           \\ 
\hline
mean of interval of arrival time of forward traffic                   &            &            &            &            & \checkmark           \\ 
\hline
mean of interval of arrival time of backward traffic                  &            &            &            &            & \checkmark           \\ 
\hline
minimum of interval of arrival time of forward traffic                &            &            &            &            & \checkmark           \\ 
\hline
length of IP packets                                                  &            &            & \checkmark          &            &             \\ 
\hline
length of TCP payload                                                 &            &            & \checkmark          &            &             \\ 
\hline
payload Ratio                                                         &            &            & \checkmark          &            &             \\ 
\hline
ratio to previous packets in each session                             &            &            & \checkmark          &            &             \\ 
\hline
time difference between packets per session                           &            &            & \checkmark          &            &             \\ 
\hline
total length of forward payload                                       & \checkmark          &            &            & \checkmark          &             \\ 
\hline
minimum length of TCP payload                                         &            &            &            & \checkmark          &             \\ 
\hline
mean length of TCP payload                                            &            &            &            & \checkmark          &             \\ 
\hline
median length of TCP payload                                          &            &            &            & \checkmark          &             \\ 
\hline
STD of the length of IP packets                        & \checkmark          &            &            & \checkmark          &             \\ 
\hline
IPratio (maximum length of IP packets / minimum length of IP packets) &            &            &            & \checkmark          &             \\ 
\hline
goodput (Total length of IP packet per session / flow duration)       &            &            &            & \checkmark          &             \\ 
\hline
maximum time difference between packets per session                   & \checkmark          &            &            &            &             \\ 
\hline
STD of forward packet length                           & \checkmark          &            &            &            &             \\ 
\hline
maximum length of TCP payload                                         & \checkmark          &            &            &            &             \\ 
\hline
mean time to live                                                     & \checkmark          &            &            &            &             \\ 
\hline
STD of time to live                                    & \checkmark          &            &            &            &             \\ 
\hline
time duration of forward traffic                                      & \checkmark          &            &            &            &             \\
\hline
\multicolumn{6}{l} {1 refers FOS feature set; 2 refers top 10 ranked features feature set; 3 refers side-channel feature }\\
\multicolumn{6}{l} {set; 4 refers tamper resistant feature set; 5 refer to time based feature set; '\checkmark' refers the feature in}\\
\multicolumn{6}{l} {this row is selected to the feature set in this column}\\
\end{tabular}
\end{table*}

For TLS/SSL features, we also use Zeek IDS to extract log files and extract features based on these log files. Table V shows some samples of the TLS/SSL features that can be extracted from different log files.

\begin{table}
\centering
\caption{Example of TLS/SSL features}
\small
\begin{tabular*}{17.5pc}{@{}|p{180pt}|@{}}
\hline 
\bf TLS/SSL features\\
\hline 
payload bytes from clients\\
\hline 
ratio of responder to clients\\
\hline 
no. of certificate\\
\hline 
TLS version types\\
\hline 
SNI in SAN DNS\\
\hline 
differ SNI in SSL log\\
\hline 
differ Subject in SSL log\\
\hline 
mean of certificate validity\\
\hline 
Common Name in SAN DNS\\
\hline 
no of domain in SAN\\
\hline 
.........\\
\hline 

\multicolumn{1}{@{}p{17.5pc}@{}}{}\\
\end{tabular*}
\end{table}

Similar to numerical feature selection, we conducted a statistical analysis on TLS/SSL features as well. A list of TLS/SSL features that appeared at high frequencies in many works is created. The list contains 22 features, such as \textit{Server Name Indication (SNI) in SAN Domain Name System (DNS)}, \textit{mean of public certificate key}, \textit{average domain in SAN}, and \textit{STD of public certificate key}. We name this list of TLS/SSL features as the Further Optimized TLS/SSL (FOTS) feature set. Since such protocol-specific features are only applicable to their related encrypted protocols, they are limited in encrypted malicious traffic detection and classification area. Therefore, this FOTS feature set is used to conduct the comparative experiment with the protocol-agnostic numerical feature set in Experiment 2 to provide some insights on which feature set is better in performance, and how well one fares against the other. We will analyze the potential of replacing TLS/SSL features (protocol-specific features) with protocol-agnostic numerical features.

\begin{table*}
\centering
\caption{Performance results of 10 algorithms with 5 different feature sets}
\begin{tabular}{|l|l|l|l|l|l|l|l|l|} 
\hline
                     & \multicolumn{8}{l|}{\textbf{FOS Feature Set}}                                                                      \\ 
\hline
                     & accuracy             & STD(accuracy) & roc-auc         & STD(roc-auc) & FPR             & STD(FPR) & TPR             & STD(TPR)  \\ 
\hline
\textbf{RF}          & \textbf{0.9471} & 0.0016   & \textbf{0.9909} & 0.0004   & \textbf{0.0418} & 0.0023   & \textbf{0.9354} & 0.0019    \\ 
\hline
\textbf{XGBoost}     & 0.9365          & 0.0011   & 0.9876          & 0.0002   & 0.0547          & 0.0027   & 0.9272          & 0.0033    \\ 
\hline
\textbf{C4.5}        & 0.9345          & 0.0039   & 0.9345          & 0.0039   & 0.0639          & 0.0040   & 0.9328          & 0.0044    \\ 
\hline
\textbf{AdaBoost}    & 0.9316          & 0.0029   & 0.9316          & 0.0028   & 0.0673          & 0.0042   & 0.9305          & 0.0026    \\ 
\hline
\textbf{CART}        & 0.9322          & 0.0029   & 0.9322          & 0.0028   & 0.0665          & 0.0038   & 0.9308          & 0.0030    \\ 
\hline
\textbf{KNN}         & 0.8651          & 0.0014   & 0.8651          & 0.0014   & 0.1315          & 0.0018   & 0.8615          & 0.0027    \\ 
\hline
\textbf{MLP}         & 0.7681          & 0.0556   & 0.8383          & 0.0751   & 0.1327          & 0.1183   & 0.6629          & 0.2145    \\ 
\hline
\textbf{Naïve Bayes} & 0.7397          & 0.0013   & 0.7990          & 0.0015   & 0.1620          & 0.0038   & 0.6355          & 0.0038    \\ 
\hline
\textbf{Logistc R}   & 0.7210          & 0.0077   & 0.7924          & 0.0016   & 0.2062          & 0.0146   & 0.6440          & 0.0018    \\ 
\hline
\textbf{Linear R}    & 0.6986          & 0.0526   & 0.6986          & 0.0507   & 0.3035          & 0.1187   & 0.7008          & 0.0183    \\ 
\hline
                     & \multicolumn{8}{l|}{\textbf{Top 10 Ranked Features Feature Set}}                                                   \\ 
\hline
                     & accuracy             & STD(accuracy) & roc-auc         & STD(roc-auc) & FPR             & STD(FPR) & TPR             & STD(TPR)  \\ 
\hline
\textbf{RF}          & \textbf{0.9400} & 0.0010   & \textbf{0.9879} & 0.0005   & \textbf{0.0571} & 0.0024   & \textbf{0.9369} & 0.0036    \\ 
\hline
\textbf{XGBoost}     & 0.9166          & 0.0020   & 0.9783          & 0.0008   & 0.0853          & 0.0044   & 0.9185          & 0.0029    \\ 
\hline
\textbf{C4.5}        & 0.9164          & 0.0031   & 0.9164          & 0.0031   & 0.0831          & 0.0036   & 0.9159          & 0.0042    \\ 
\hline
\textbf{AdaBoost}    & 0.9153          & 0.0027   & 0.9153          & 0.0027   & 0.0836          & 0.0025   & 0.9141          & 0.0033    \\ 
\hline
\textbf{CART}        & 0.9151          & 0.0023   & 0.9151          & 0.0023   & 0.0838          & 0.0020   & 0.9140          & 0.0034    \\ 
\hline
\textbf{KNN}         & 0.8566          & 0.0035   & 0.8566          & 0.0035   & 0.1404          & 0.0045   & 0.8534          & 0.0030    \\ 
\hline
\textbf{MLP}         & 0.7571          & 0.0594   & 0.7924          & 0.0786   & 0.0744          & 0.0743   & 0.5786          & 0.1830    \\ 
\hline
\textbf{Naïve Bayes} & 0.7580          & 0.0018   & 0.8068          & 0.0040   & 0.1391          & 0.0031   & 0.6490          & 0.0035    \\ 
\hline
\textbf{Logistc R}   & 0.7430          & 0.0018   & 0.8026          & 0.0016   & 0.2067          & 0.0174   & 0.6897          & 0.0192    \\ 
\hline
\textbf{Linear R}    & 0.5543          & 0.0684   & 0.5591          & 0.0688   & 0.6080          & 0.1469   & 0.7262          & 0.1600    \\ 
\hline
                     & \multicolumn{8}{l|}{\textbf{Side Channel Feature Set}}                                                             \\ 
\hline
                     & accuracy             & STD(accuracy) & roc-auc         & STD(roc-auc) & FPR             & STD(FPR) & TPR             & STD(TPR)  \\ 
\hline
\textbf{RF}          & 0.8190          & 0.0006   & 0.9147          & 0.0004   & 0.1788          & 0.0017   & \textbf{0.8166} & 0.0011    \\ 
\hline
\textbf{XGBoost}     & 0.8235          & 0.0018   & \textbf{0.9181} & 0.0012   & 0.1631          & 0.0014   & 0.8093          & 0.0038    \\ 
\hline
\textbf{C4.5}        & 0.8128          & 0.0007   & 0.8517          & 0.0005   & 0.1764          & 0.0015   & 0.8013          & 0.0011    \\ 
\hline
\textbf{AdaBoost}    & 0.8158          & 0.0005   & 0.8711          & 0.0003   & 0.1759          & 0.0100   & 0.8070          & 0.0109    \\ 
\hline
\textbf{CART}        & 0.8126          & 0.0006   & 0.8515          & 0.0005   & 0.1767          & 0.0014   & 0.8013          & 0.0014    \\ 
\hline
\textbf{KNN}         & \textbf{0.8259} & 0.0032   & 0.9124          & 0.0019   & 0.1279          & 0.0073   & 0.7770          & 0.0132    \\ 
\hline
\textbf{MLP}         & 0.6717          & 0.0107   & 0.7477          & 0.0088   & 0.1373          & 0.0825   & 0.4694          & 0.0926    \\ 
\hline
\textbf{Naïve Bayes} & 0.6244          & 0.0006   & 0.6135          & 0.0012   & \textbf{0.0223} & 0.0007   & 0.2503          & 0.0009    \\ 
\hline
\textbf{Logistc R}   & 0.5973          & 0.0010   & 0.6417          & 0.0008   & 0.2900          & 0.0018   & 0.4779          & 0.0020    \\ 
\hline
\textbf{Linear R}    & 0.5649          & 0.0427   & 0.5684          & 0.0360   & 0.5558          & 0.2697   & 0.6926          & 0.1980    \\ 
\hline
                     & \multicolumn{8}{l|}{\textbf{Tamper Resistant Feature Set}}                                                         \\ 
\hline
                     & accuracy             & STD(accuracy) & roc-auc         & STD(roc-auc) & FPR             & STD(FPR) & TPR             & STD(TPR)  \\ 
\hline
\textbf{RF}          & 0.8828          & 0.0013   & 0.9616          & 0.0005   & \textbf{0.1100} & 0.0018   & 0.8753          & 0.0029    \\ 
\hline
\textbf{XGBoost}     & \textbf{0.8894} & 0.0010   & \textbf{0.9654} & 0.0008   & 0.1186          & 0.0037   & 0.8978          & 0.0045    \\ 
\hline
\textbf{C4.5}        & 0.8790          & 0.0031   & 0.8790          & 0.0031   & 0.1189          & 0.0051   & 0.8768          & 0.0016    \\ 
\hline
\textbf{AdaBoost}    & 0.8785          & 0.0033   & 0.9111          & 0.0329   & 0.1189          & 0.0055   & 0.8758          & 0.0027    \\ 
\hline
\textbf{CART}        & 0.8762          & 0.0019   & 0.8762          & 0.0019   & 0.1213          & 0.0041   & 0.8737          & 0.0013    \\ 
\hline
\textbf{KNN}         & 0.8469          & 0.0022   & 0.8471          & 0.0023   & 0.1529          & 0.0016   & 0.8467          & 0.0039    \\ 
\hline
\textbf{MLP}         & 0.7510          & 0.1026   & 0.8547          & 0.0581   & 0.2858          & 0.2939   & 0.7899          & 0.1360    \\ 
\hline
\textbf{Naïve Bayes} & 0.4782          & 0.0015   & 0.7353          & 0.0021   & 0.9825          & 0.0019   & \textbf{0.9661} & 0.0018    \\ 
\hline
\textbf{Logistc R}   & 0.6591          & 0.0040   & 0.7580          & 0.0043   & 0.4832          & 0.0051   & 0.8097          & 0.0057    \\ 
\hline
\textbf{Linear R}    & 0.7482          & 0.0061   & 0.7442          & 0.0063   & 0.1194          & 0.0552   & 0.6079          & 0.0583    \\ 
\hline
                     & \multicolumn{8}{l|}{\textbf{Time Based Feature Set}}                                                               \\ 
\hline
                     & accuracy             & STD(accuracy) & roc-auc         & STD(roc-auc) & FPR             & STD(FPR) & TPR             & STD(TPR)  \\ 
\hline
\textbf{RF}          & \textbf{0.9207} & 0.0010   & \textbf{0.9792} & 0.0006   & \textbf{0.0759} & 0.0010   & \textbf{0.9171} & 0.0022    \\ 
\hline
\textbf{XGBoost}     & 0.9124          & 0.0033   & 0.9746          & 0.0012   & 0.0831          & 0.0047   & 0.9076          & 0.0048    \\ 
\hline
\textbf{C4.5}        & 0.9119          & 0.0030   & 0.9119          & 0.0030   & 0.0869          & 0.0037   & 0.9107          & 0.0035    \\ 
\hline
\textbf{AdaBoost}    & 0.9136          & 0.0014   & 0.9135          & 0.0014   & 0.0856          & 0.0014   & 0.9127          & 0.0028    \\ 
\hline
\textbf{CART}        & 0.9139          & 0.0023   & 0.9138          & 0.0023   & 0.0849          & 0.0023   & 0.9126          & 0.0037    \\ 
\hline
\textbf{KNN}         & 0.8667          & 0.0021   & 0.8678          & 0.0019   & 0.1355          & 0.0024   & 0.8691          & 0.0023    \\ 
\hline
\textbf{MLP}         & 0.8056          & 0.0091   & 0.9025          & 0.0049   & 0.2023          & 0.0620   & 0.8140          & 0.0488    \\ 
\hline
\textbf{Naïve Bayes} & 0.6339          & 0.0019   & 0.5832          & 0.0015   & 0.1038          & 0.0012   & 0.3561          & 0.0045    \\ 
\hline
\textbf{Logistc R}   & 0.6557          & 0.0010   & 0.6918          & 0.0036   & 0.1204          & 0.0012   & 0.4186          & 0.0025    \\ 
\hline
\textbf{Linear R}    & 0.5799          & 0.0750   & 0.5830          & 0.0754   & 0.5288          & 0.3559   & 0.6948          & 0.3713    \\
\hline
\end{tabular}
\end{table*}

\subsection{\bf Experiment Setup}

Ten commonly used machine learning algorithms are chosen in our experiment. They are RF, KNN, CART, C4.5, MLP, Naïve Bayes, XGBoost, AdaBoost, Linear Regression (Linear R), and Logistic Regression (Logistic R). In order to construct our models, scikit-learn and xgboost libraries for Python are used with default hyperparameters (except RF's n\_estimators = 200 and KNN's n\_neighbors = 6). The stratified cross validation is used for model training based on k-fold, where k is set to 5. 

In recent years, many authors have stated in their papers that deep learning is becoming a more efficient technology in the research of encrypted malicious traffic detection, while traditional machine learning approaches have not been considered, or simply discussed without in-depth analysis. We agree that the proportion of research papers on deep learning encrypted malicious traffic detection is increasing, and deep learning is indeed gradually replacing traditional machine learning as a broader technology for detecting encrypted malicious traffic. However, this does not mean that traditional machine learning has lost the value of research and commentary. Therefore, in this paper, we not only conduct a comprehensive review of recent deep learning and machine learning papers' contributions, but also conduct comparative experiments on commonly used traditional machine learning. Furthermore, one of our experiment purposes is to deliver a more optimal feature set. Machine learning algorithms which can be used to effectively evaluate optimal feature sets in a straightforward manner.

The experiment running: Intel(R) Core(TM) i7-10700K CPU @ 3.8GHz 64.0GB of RAM. Experiments evaluated ROC-AUC, accuracy, FPR, and true positive rate(TPR) values of detection models. Furthermore, the STD information of each evaluation measure is also calculated.

\subsection{\bf Experiment 1: Performance Analysis of the different statistical numerical feature sets and algorithms using a mixed dataset}

We would like to achieve four experiment objectives by conducting this experiment: 

1. Analyze the performance of different algorithms using the same feature set. 

2. Analyze the performance of different feature sets using the same algorithms. 

3. Identify the optimal algorithm and feature set using this dataset.

4. Cross Dataset Validation to highlight the importance of using the composed dataset instead of using datasets for model training separately.

Therefore, in the first experiment, we first utilized the composed dataset to train ten different algorithms with five different protocol-agnostic numerical feature sets mentioned in section V.B. The experiment results are shown from Fig. 4 to Fig. 7, and Table VI.

\begin{figure}
\centerline{\includegraphics[width=21pc]{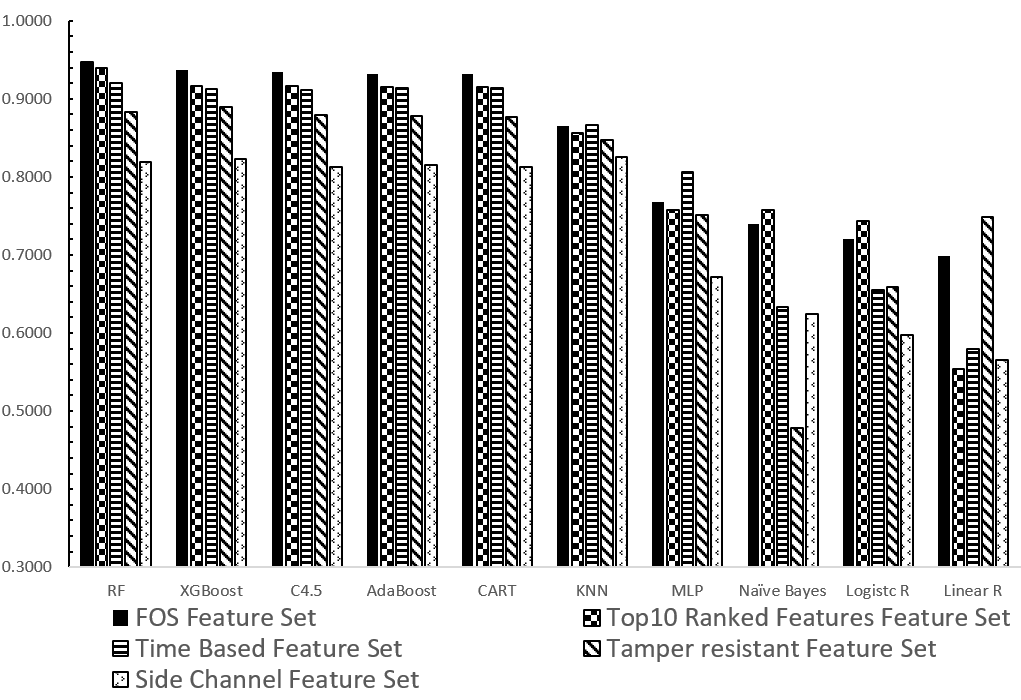}}
\caption{Accuracy performance of 10 algorithms with 5 different feature sets}
\end{figure}

\begin{figure}
\centerline{\includegraphics[width=21pc]{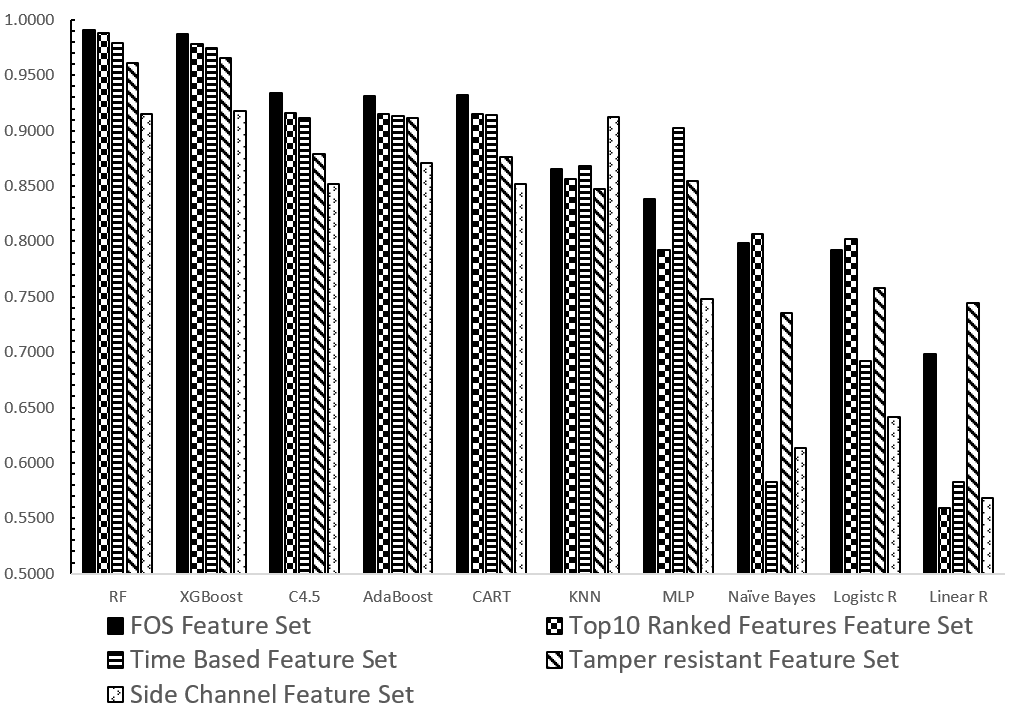}}
\caption{ROC-AUC performance of 10 algorithms with 5 different feature sets}
\end{figure}

\begin{figure}
\centerline{\includegraphics[width=21pc]{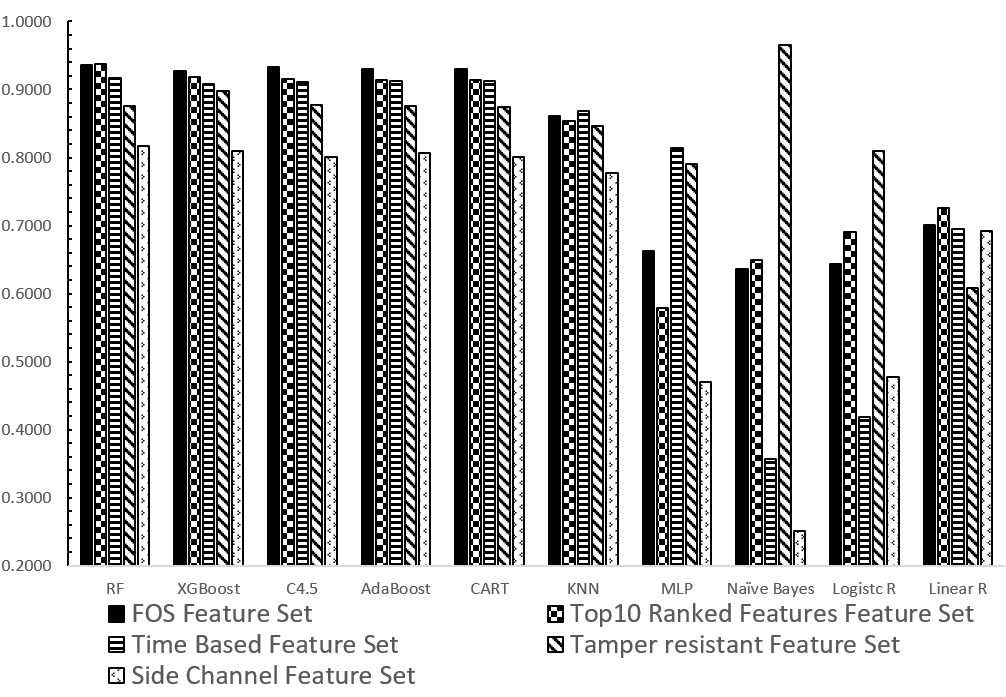}}
\caption{TPR performance of 10 algorithms with 5 different feature sets}
\end{figure}

\begin{figure}
\centerline{\includegraphics[width=21pc]{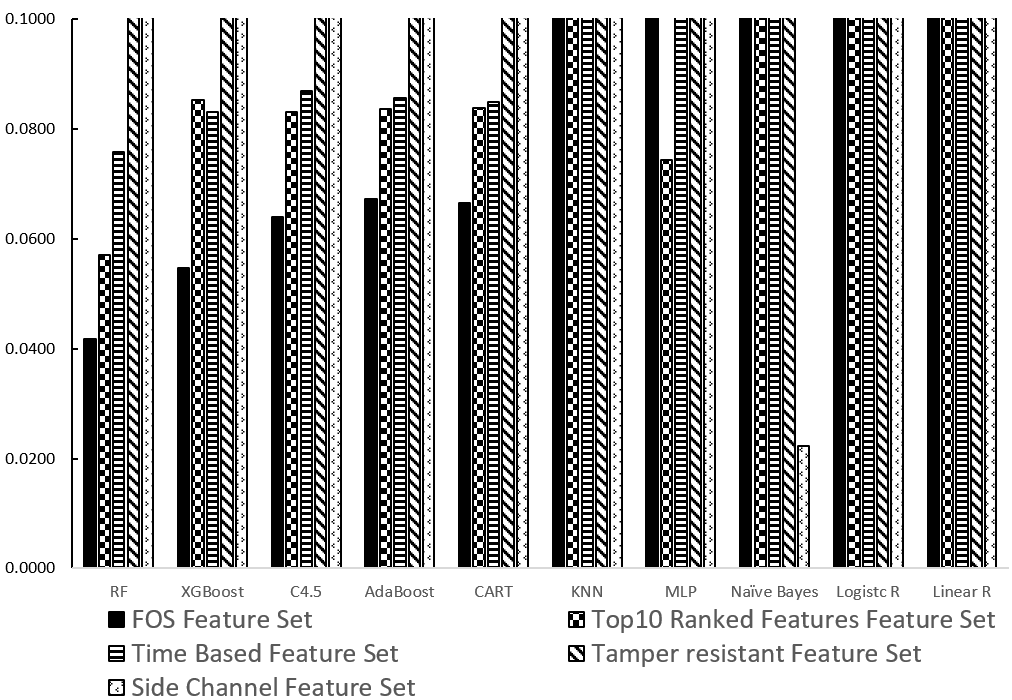}}
\caption{FPR performance of 10 algorithms with 5 different feature sets}
\end{figure}

\begin{figure}
\centerline{\includegraphics[width=21pc]{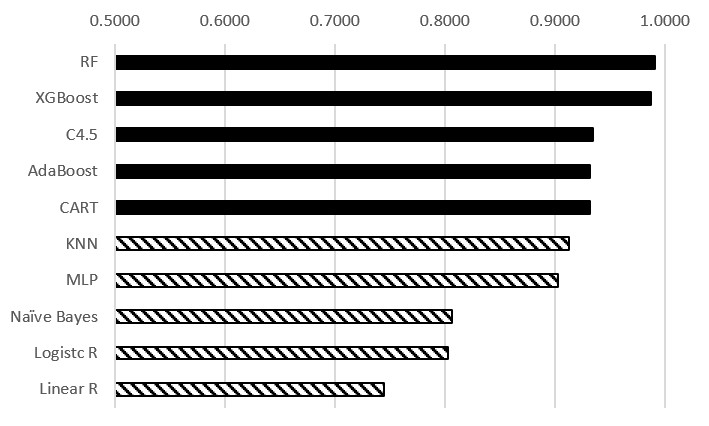}}
\caption{The highest ROC-AUC among five feature sets under each algorithms. Black solid fill bars represent algorithms with FOS feature set and other bars represent algorithms with other feature sets.}
\end{figure}

\begin{figure}
\centerline{\includegraphics[width=21pc]{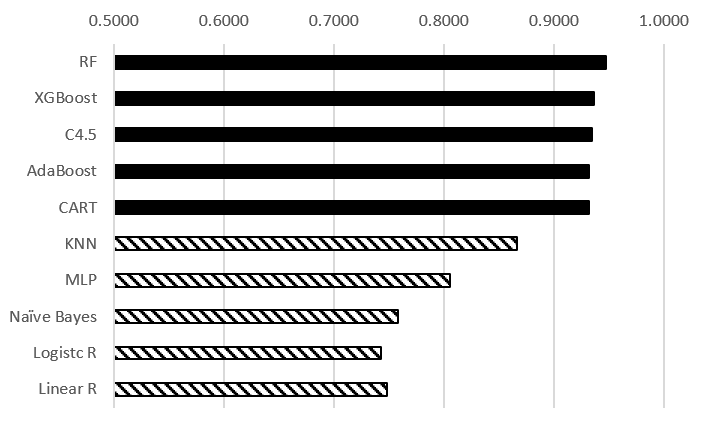}}
\caption{The highest accuracy among five feature sets under each algorithms. Black solid fill bars represent algorithms with FOS feature set and other bars represent algorithms with other feature sets.}
\end{figure}

\begin{figure}
\centerline{\includegraphics[width=21pc]{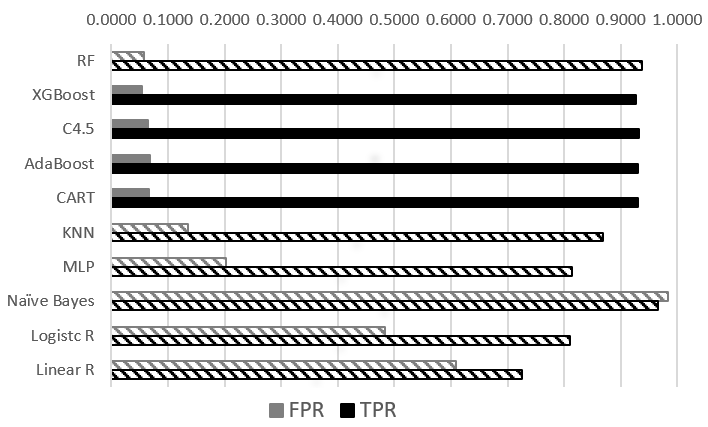}}
\caption{The highest TPR among five feature sets under each algorithms and their corresponding FPR. Black solid fill bars represent algorithms with FOS feature set and other bars represent algorithms with other feature sets.}
\end{figure}

\begin{figure}
\centerline{\includegraphics[width=21pc]{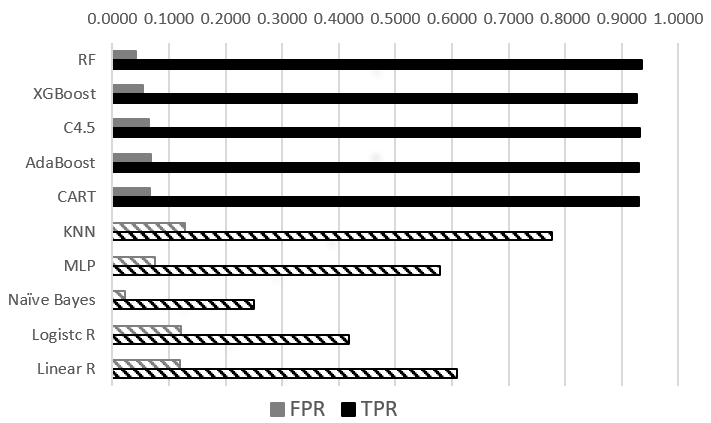}}
\caption{The lowest FPR among five feature sets under each algorithms and their corresponding TPR. Black solid fill bars represent algorithms with FOS feature set and other bars represent algorithms with other feature sets.}
\end{figure}

For the experiment’s first objective, we can refer to Fig. 4 to Fig. 7 and Table VI for further analysis. In Fig. 4, RF, XGBoost, C4.5, AdaBoost, CART, and KNN outperform the remaining algorithms. We can observe that the accuracy is above 80\% and the STD of accuracy is lower than 0.5\% for all feature sets with above 6 algorithms. In Fig. 5, ROC-AUC values of both RF and XGBoost outperform other algorithms. Regardless of which feature set is used for RF and XGBoost model training, their ROC-AUC can be higher than 90\% and STD of ROC-AUC can be lower than 0.13\%, which is not possible with other algorithms in our experiment. Next is to compare Fig. 6 and Fig. 7 (TPR VS. FPR). Based on Fig. 6, Fig. 7, and Table VI, tamper resistant feature set with Naïve Bayes achieved the highest TPR (96.61\%) in this experiment, but its FPR is high which is 98.25\%. Side channel feature with Naïve Bayes achieved the lowest FPR (2.23\%) in this experiment, but its TPR is only 25.03\%. Therefore, we cannot accept the FPR of side channel feature set with Naïve Bayes and TPR of tamper resistant feature set with Naïve Bayes. RF and XGBoost models that have always performed well in Fig. 4 and Fig. 5 can also achieve low FPR and high TPR. FOS feature set with either RF or XGBoost can simultaneously achieve TPR higher than 92\% and FPR lower than 6\%. Compared with RF and XGBoost, TPR and FPR of RF are better than XGBoost. In particular, the RF model with FOS feature set is able to achieve 94.71\% accuracy, 99.09\% ROC-AUC and 4.18\% FPR, which outperforms all other algorithms. Its 93.54\% TPR also achieved the second highest value (The highest is 93.69\% for RF with Top 10 ranked features feature set).

For the experiment's second objective, the analysis outcomes of the comparison are summarized and plotted in Fig. 8 to Fig. 11. These figures represent the best performance of each algorithm among five feature sets in terms of accuracy, ROC-AUC, TPR, and FPR. From Fig. 8 to Fig. 9, a solid bar indicates the best performance of this algorithm is achieved by FOS feature set. A bar filled with diagonal lines indicates the best performance of this algorithm is achieved by the other four feature sets. Based on Fig. 8 and Fig. 9, we observed that FOS feature set with RF, XGBoost, C4.5, AdaBoost, and CART can achieve the highest ROC-AUC and accuracy values among all feature sets. In KNN, MLP, Naïve Bayes, Logistic R, and Linear R models, the highest ROC-AUC and accuracy are achieved by the other four feature sets, but their performances are all lower than the first five algorithms with FOS feature set.

In Fig. 10 and Fig. 11, grey color indicates FPR and black color indicates TPR. Furthermore, a solid filled bar indicates the best performance of FPR or TPR of this algorithm is achieved by FOS feature set. A bar filled with diagonal lines indicates the best performance of FPR or TPR of this algorithm is achieved by the other four feature sets. For Fig. 11, we can observe that the lowest FPR of five algorithms are achieved by FOS feature set, and their corresponding TPRs are all greater than 92\%. Among them, FOS feature set with RF achieved the lowest FPR (The lowest FPR of Naïve Bayes is not acceptable, which has been discussed above). For Fig. 10, four algorithms with FOS feature set achieved the highest TPR, of which XGBoost, C4.5, AdaBoost, and CART with FOS feature set achieved both highest TPR and Lowest FPR, and all have achieved an STD of lower than 0.5\%. However, the highest TPR of RF is not achieved by FOS feature set, but Top 10 ranked features feature set. In order to find the optimal feature set between FOS feature set and Top 10 ranked features feature set, we also need to consider the exact performance results in Table VI. Based on Table VI, FOS feature set with RF achieved 93.54\% TPR and 4.18\% FPR, and Top 10 ranked features feature set with RF achieved 93.69\% TPR and 5.71\% FPR. Although the TPR of FOS feature set with RF is 0.15\% lower than that of Top 10 ranked features feature set with RF, the FPR of FOS feature set with RF is 1.53\% lower than that of Top 10 ranked features feature set with RF. Therefore, FOS feature set with RF provides a better trade off between TPR and FPR. 

From the analysis results based on the experiment's first and second objectives, it is apparent that the combination of the RF algorithm and FOS feature set can achieve the best performance.

Next, we will conduct a more detailed analysis of this experiment. The experiments showed that a feature set that combined time based features and traffic numerical features can enhance the model detection rate as compared to using either one alone. For instance, FOS feature set and Top 10 ranked features feature set both contain time based features and traffic numerical features, but time based feature set contains the time based features only. Furthermore, in the experiment, side channel feature set has the worst performance among the five feature sets. We believe that it is because the features in the side channel feature set are all packet based features, such as payload size, packet size, and payload ratio. Such features may not have a clear distinction in values between malicious and legitimate traffic flows, which may affect the performance of the detection models. While the features of the other four feature sets are session based features, they consider multiple packets sequence instead of only single packets. The experiment indicates that session based features or a combination of session based features with packet based features are likely to outperform those that consider packet based features only.

\begin{table*}
\centering
\caption{Cross dataset validation results based on RF with FOS feature set}
\begin{tabular}{|l|l|l|l|l|l|} 
\hline
\begin{tabular}[c]{@{}l@{}}\textbf{Train Other Datasets }\\\textbf{and Test Dataset A}\end{tabular} & \begin{tabular}[c]{@{}l@{}}\textbf{A=}\\\textbf{CICIDS-2012}\end{tabular} & \begin{tabular}[c]{@{}l@{}}\textbf{A=}\\\textbf{CICIDS-2017}\end{tabular} & \begin{tabular}[c]{@{}l@{}}\textbf{\textbf{A=}}\\\textbf{\textbf{Malware Capture~}}\\\textbf{\textbf{Facility Project~Dataset}}\end{tabular}                                       & \begin{tabular}[c]{@{}l@{}}\textbf{A=}\\\textbf{UNSW NS2019 }\\\textbf{Dataset}\end{tabular}                           & \begin{tabular}[c]{@{}l@{}}\textbf{A=}\\\textbf{CIC-AndMal2017}\end{tabular}             \\ 
\hline
Accuracy                                                                                            & 0.2868                                                                    & 0.8005                                                                    & 0.4318                                                                                                                                       & 0.4873                                                                                                                 & 0.5840                                                                                   \\ 
\hline
roc\_auc                                                                                            & 0.3026                                                                    & 0.8148                                                                    & 0.7451                                                                                                                                       & 0.5830                                                                                                                 & 0.6222                                                                                   \\ 
\hline
FPR                                                                                                 & 0.6970                                                                    & 0.3799                                                                    & 0.0881                                                                                                                                       & 0.0507                                                                                                                 & 0.0002                                                                                   \\ 
\hline
TPR                                                                                                 & 0.2702                                                                    & 0.9878                                                                    & 0.0018                                                                                                                                       & 0.0110                                                                                                                 & 0.0018                                                                                   \\ 
\hline
\begin{tabular}[c]{@{}l@{}}\textbf{Train Dataset A and }\\\textbf{Test Other Datasets}\end{tabular} & \begin{tabular}[c]{@{}l@{}}\textbf{A=}\\\textbf{CICIDS-2012}\end{tabular} & \begin{tabular}[c]{@{}l@{}}\textbf{A=}\\\textbf{CICIDS-2017}\end{tabular} & \begin{tabular}[c]{@{}l@{}}\textbf{\textbf{A=}}\\\textbf{\textbf{Malware Capture~}}\\\textbf{\textbf{Facility Project~Dataset}}\end{tabular} & \begin{tabular}[c]{@{}l@{}}\textbf{\textbf{A=}}\\\textbf{\textbf{UNSW NS2019}}\\\textbf{\textbf{Dataset}}\end{tabular} & \begin{tabular}[c]{@{}l@{}}\textbf{A=}\\\textbf{CIC-AndMal2017}\end{tabular}  \\ 
\hline
Accuracy                                                                                            & 0.5422                                                                    & 0.5190                                                                    & 0.6528                                                                                                                                       & 0.5165                                                                                                                 & 0.4906                                                                                   \\ 
\hline
roc\_auc                                                                                            & 0.6450                                                                    & 0.6552                                                                    & 0.7030                                                                                                                                       & 0.7783                                                                                                                 & 0.4705                                                                                   \\ 
\hline
FPR                                                                                                 & 0.2260                                                                    & 0.0041                                                                    & 0.2500                                                                                                                                       & 0.0000                                                                                                                 & 0.5283                                                                                   \\ 
\hline
TPR                                                                                                 & 0.2960                                                                    & 0.0107                                                                    & 0.5445                                                                                                                                       & 0.0000                                                                                                                 & 0.5093                                                                                   \\
\hline
\end{tabular}
\end{table*}

Furthermore, a cross dataset validation is also conducted in this experiment. The purpose of this is to highlight the importance of using the composed dataset instead of using single datasets for model training and validation. We perform two such experiments: Train selected Dataset A and test the other 4 datasets; and train 4 datasets and test Dataset A. we select the RF with FOS feature set for both above cross dataset validation tasks, because it reported the best performance. The experiment results are shown in Table VII.

By observing the two cross dataset validation results in Table VII, we obtained very poor accuracy, ROC-AUC, FPR, and TPR for both tasks. Both FPR and TPR are very low, which indicates that the model classifies most of the test set data as negative. Such results may be expected that their data structure is very different in distribution, especially the difference of data distribution between IoT and conventional device data. When we use the conventional device data to train the RF model with FOS feature set and use the IoT dataset as the test set, we get 0 FPR and 0 TPR at the same time. This shows that in the experiment, the RF model trained by the conventional device data has almost no detection capabilities for the IoT test set. Therefore, a single public dataset would have a lack of variety in encryption traffic. Such limitations may lead to certain unpredictable biases, such as the high possibility of over-fitting in the selected dataset. In summary, the above analysis highlights the importance of using a comprehensively composed dataset instead of using datasets for model training separately.

\subsection{\bf Experiment 2: Compare the performance of Protocol-agnostic numerical and TLS/SSL feature sets}

In the second experiment, the aim is to evaluate and compare the performance of the protocol-agnostic numerical features and TLS/SSL features. As TLS/SSL features are only limited to traffic encrypted by these protocols, we use a dataset containing only TLS/SSL encrypted traffic for this experiment. We applied FOTS feature set and FOS feature set to train the models using the ten different algorithms. The experiment result is shown in Fig. 12 and Fig. 13.

\begin{figure}
\centerline{\includegraphics[width=21pc]{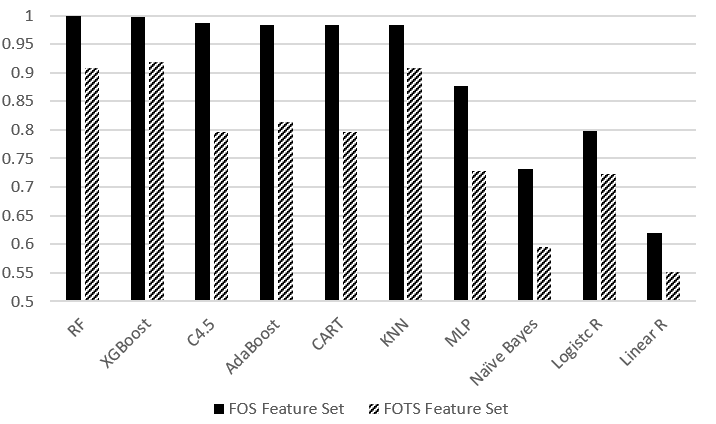}}
\caption{ROC-AUC result of Experiment 2 between statistical numerical and TLS/SSL features}
\end{figure}

\begin{figure}
\centerline{\includegraphics[width=21pc]{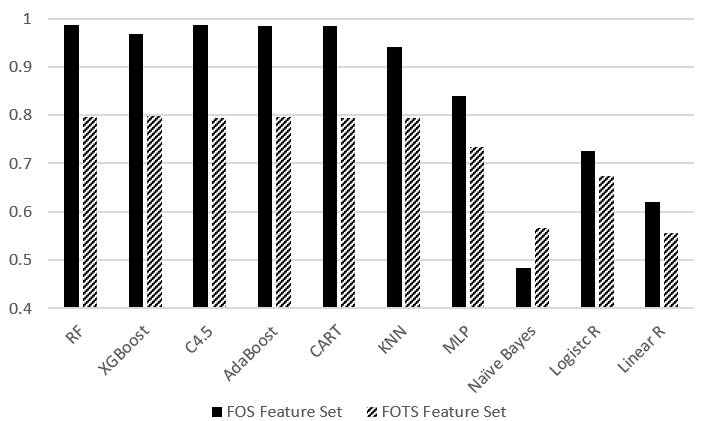}}
\caption{Accuracy result of Experiment 2 between statistical numerical and TLS/SSL features}
\end{figure}

Based on Fig. 12 and Fig. 13, we can observe that except for Naïve Bayes, the performance of the other nine algorithms with FOS feature set can achieve a higher or similar performance in terms of both accuracy and ROC-AUC compare with using the TLS/SSL features under the same condition. While only the Naïve Bayes with FOS feature set achieved higher ROC-AUC, but lower accuracy than the Naïve Bayes using TLS/SSL features. Although, there is a different performance of Naïve Bayes, the performance of Naïve Bayes detection model is lower than most other algorithms no matter FOS feature set or TLS/SSL features is selected. In summary, Among top performance algorithms, such as RF and XGBoost, FOS feature set can achieve better performance than using the TLS/SSL features regardless of using ROC-AUC or accuracy as a performance evaluation measure. Therefore, it is feasible to consider replacing protocol-specific features with protocol-agnostic numerical features. Then, most research on encrypted malicious traffic detection will no longer be limited to HTTPs or any specific protocols, but a more comprehensive encrypted malicious traffic analysis can be carried out. The future research of feature selection and optimization on protocol-agnostic numerical features may be more meaningful than protocol-specific features as well.

\section{Conclusion}

In the paper, we proposed a framework to study and analyse the machine learning based encrypted traffic detection approach. We reviewed existing research based on the proposed framework, including the research objective construction, traffic dataset collection and pre-processing, feature extraction and selection, algorithm selection and performance evaluation.

While some progress has been made in encrypted malicious traffic detection, challenges still exist. The first and most important issue to be addressed is the lack of a comprehensive, class balanced, realistic and convincing public dataset in the area of encrypted malicious traffic detection. The quality of the dataset used by research has a direct impact on the final performance of its model. Thus, we analysed, processed and combined 5 publicly available datasets to construct a large and comprehensive dataset to facilitate future research and technology evaluation in this field.

In addition, we trained detection models using 10 machine learning algorithms and 5 feature sets, and conducted comparative experiments to analyse the performance of different protocol-agnostic numerical feature sets and algorithms using the same dataset. We also tested the feasibility of replacing TLS/SSL features with protocol-agnostic numerical features for future encrypted malicious traffic analysis.

Lastly, we also showed that the study on only one or two kinds of encrypted traffic should be avoided. With the growth in the types of encryption protocols and proprietary protocols, the analysis of one or two kinds of encryption protocols is bound to play a small role moving forward. Furthermore, current approaches are mostly binary classification, and multi-class classification can be explored to obtain more fine-grained detection results.

\end{document}